\documentclass[preprint,prb,a4paper,floats,titlepage,fleqn,footinbib,lengthcheck,twocolumn,tightenlines,balancelastpage,10pt,superscriptaddress,showpacs]{revtex4-1}
\usepackage{natbib}
\usepackage{amssymb}
\usepackage{amsfonts}
\usepackage{amsmath}
\usepackage{graphicx}
\usepackage{color}
\usepackage{longtable}
\usepackage{revsymb}
\usepackage{units}
\usepackage{nicefrac}
\usepackage{textcomp}
\usepackage{epsfig}
\usepackage{footmisc}
\usepackage{soul}
\usepackage{ulem}

\ifx\pdfoutput\relax\let\pdfoutput=\undefined\fi
\newcount\msipdfoutput
\ifx\pdfoutput\undefined\else
\ifcase\pdfoutput\else
\ifx\paperwidth\undefined\else
\ifdim\paperheight=0pt\relax\else\pdfpageheight\paperheight\fi
\ifdim\paperwidth=0pt\relax\else\pdfpagewidth\paperwidth\fi
\fi\fi\fi
\begin{document}

\title{On the bad metallicity and phase diagrams of Fe$_{1+\delta }X$ ($X$%
=Te, Se, S, solid solutions): an electrical resistivity study}
\author{M. ElMassalami}
\affiliation{Instituto de F\'{\i}sica, Universidade Federal do Rio de Janeiro, Caixa
Postal 68528, 21941-972 Rio de Janeiro RJ, Brazil}
\author{K. Deguchi}
\affiliation{National Institute for Materials Science, 1-2-1, Sengen, Tsukuba, 305-0047,
Japan}
\author{T. Machida}
\affiliation{Department of Physics, Tokyo University of Science, 1-3 Kagurazaka,
Shinjuku-ku, Tokyo 162-8601, Japan}
\author{H. Takeya }
\affiliation{National Institute for Materials Science, 1-2-1, Sengen, Tsukuba, 305-0047,
Japan}
\author{Y. Takano}
\affiliation{National Institute for Materials Science, 1-2-1, Sengen, Tsukuba, 305-0047,
Japan}

\begin{abstract}
Based on a systematic analysis of the thermal evolution of the resistivities
of Fe-based chalcogenides Fe$_{1+\delta }$Te$_{1-x}X_{x}$ ($X$= Se, S), it
is inferred that their often observed nonmetallic resistivities are related
to a presence of two resistive channels: one is a high-temperature
thermally-activated process while the other is a low-temperature log-in-$T$
process. On lowering temperature, there are often two metal-to-nonmetall
crossover events: one from the high-$T$ thermally-activated nonmetallic
regime into a metal-like phase and the other from the log-in-$T$ regime into
a second metal-like phase. Based on these events, together with the magnetic
and superconducting transitions, a phase diagram is constructed for each
series. We discuss the origin of both processes as well as the associated
crossover events. We also discuss how these resistive processes are being
influenced by pressure, intercalation, disorder, doping, or sample condition
and, in turn, how these modifications are shaping the associated phase
diagrams.
\end{abstract}

\date{\today}
\maketitle

\section{Introduction}

It is remarkable that the normal-state resistivities of Fe$_{1+\delta }X$ ($%
X $= Te, Se, S, or their solid solutions) \cite%
{Mizuguchi10-FeT-Review,Deguchi12-FeX-Review,
Liu10-Fe(TeSe)-PhaseDiagram,*Katayama10-Fe(SeTe)-phaseDiagram,Sales09-FeTeSe-SUC,*Tropeano10-Fe(TeSe)-transport}
as well as those of intercalated $A_{x}$Fe$_{2-y}$Se$_{2}$ ($A$=K, Rb, Cs,
Tl, ..)\cite%
{Guo10-KxFe2Se2-Sup,Fang11-(TlK)FexSe2,*Ying12-AxFe2Se2,*Wang11-doping-KxFe2-ySe2,Wang14-MIT-RbFe2Se2,*Chen11-KxFe2-ySe2}
are neither truly metallic nor truly insulating. It is also remarkable that
the normal-state and superconducting phase diagrams of these chalcogenides
are highly irreproducible and markedly different from one another. In fact,
reported resistivities of the very same stoichiometric compound do not show
the same thermal/magnetic/baric/concentration evolution, $\rho (T,H,P,x)$,
or the same crossover/transition events. It is then no surprise that these
phase diagrams (being constructed out of such characterization) manifest a
strong dependence on sample condition or history.\cite%
{Mizuguchi10-FeT-Review,Deguchi12-FeX-Review,Liu10-Fe(TeSe)-PhaseDiagram,*Katayama10-Fe(SeTe)-phaseDiagram,Sales09-FeTeSe-SUC,*Tropeano10-Fe(TeSe)-transport}

Although an earlier electronic structure calculations predicted a
low-carrier-density metallic character,\cite{Singh12-Chalcogenides-Review}
the isomorphous Fe$_{1+\delta }X$ compounds exhibit a variety of
normal-state behavior: Fe$_{1+\delta }$S is nonmetallic below $T_{\text{%
\textit{MNM}}}$ $\sim $300 K [Ref.{\onlinecite{Bertaut65}]} but $T_{\text{%
\textit{MNM}}}$ can be strongly reduced by pressure.{\cite%
{Denholme14-FeS-metallicity}} Fe$_{1+\delta }$Se is a nonmagnetic metal,
undergoes a structural phase transition at $T_{S}\sim $90 K, and
superconducts at $T_{c}\sim $8 K.\cite{McQueen09-FeSe-StrTransition}
Finally, Fe$_{1+\delta }$Te is a nonmetallic paramagnet above a magnetic and
structural transition at $T_{MS}\sim $70K while a metallic antiferromagnet
(AFM) below $T_{MS}$.\cite{Bao09-Fe(TeSe),Li09-FeTe-1stOrderTransition}

In this work we address the above mentioned bad metallicity of Fe$_{1+\delta
}X_{1-x}Y_{x}$ chalcogenides.\ Based on the analysis of their resistivities
and on the obtained phase diagrams, we identified two processes that are
responsible for their bad metallic character as well as for shaping their
phase diagrams: the first is a high-temperature (150 to 300 K)
thermally-activated process\cite%
{Yi13-OSMP-AxFe2-ySe2,*Gao14-OSMP-245Phase-HighPressure} while the other is
a low-temperature ($<$ 100 K) log-in-$T$ process.\cite%
{Altshuler83-Localization-e-e}

\section{Guidelines for analyzing $\protect\rho (T,P,x)$}

In order to rationalize the variety of functional forms of $\rho (T,P,x)$
and, in addition, so as to identify and evaluate the strength of the
involved resistive channels, let us assume, based on earlier studies,\cite%
{Yu13-OSMP-KxFe2-yK2,Craco11-Mott-KxFe2-ySe2,Yin12-powerLaw-Fe-Chalcogenides,Bascones12-OSMP-Magnetic-Fe-SUCs}
that the character of their normal-state is shaped by the combined
influences of crystalline electric field interactions, electronic
correlations, disorder, and band filling: based on the strength of these
factors, the high-temperature ($T>$100 K) normal state could be either
metallic, Mott insulator, or an intermediate orbital-selective Mott phase
(OSMP) wherein some of the Fe $3d$ orbitals are localized while the others
are itinerant.\cite%
{Yi13-OSMP-AxFe2-ySe2,Yu13-OSMP-KxFe2-yK2,Craco11-Mott-KxFe2-ySe2,Yin12-powerLaw-Fe-Chalcogenides,Bascones12-OSMP-Magnetic-Fe-SUCs}
The high-$T$ phase of most of the studied intercalated $A_{x}$Fe$_{2-y}$Se$%
_{2}$ (as well as most of Fe$_{1+\delta }X_{1-x}Y_{x}$, see below) compounds
is reported to be an OSMP.\cite%
{Yi13-OSMP-AxFe2-ySe2,Yu13-OSMP-KxFe2-yK2,Craco11-Mott-KxFe2-ySe2,Yin12-powerLaw-Fe-Chalcogenides,Bascones12-OSMP-Magnetic-Fe-SUCs}
We assumed that, within this OSMP, localized states are separated from
itinerant ones by a mobility edge at $E_{c}$.\cite{Mott87-MobilityEdge} Then
the thermal evolution of the resistivity depends on the relative strength of 
$\left\vert E_{c}-E_{F}\right\vert $ with respect to $k_{B}T$: if $E_{c}$ is
not located in the 3$d$ multiplet or that $\left\vert E_{c}-E_{F}\right\vert
>k_{B}T$, then transport is effected by a thermally-assisted hopping among
the localized orbitals leading to a Mott\ variable range hopping resistivity
(VRH):%
\begin{equation}
\rho (T)=\rho _{0}^{vrh}\exp (\left[ T_{vrh}/T\right] ^{\frac{1}{d+1}}),
\label{Eq.Mott-VRH}
\end{equation}%
where $d$=2 (3) represents a 2- (3-) dimensionality and all other terms have
their usual meaning. If $\left\vert E_{c}-E_{F}\right\vert <k_{B}T$, as
assumed for the under-study Fe-based compounds, then $\rho (T)$ is governed
by the Arrhenius process:{\cite{Basko06-MIT}} 
\begin{equation}
\rho (T)=\rho _{0}^{A}\exp (\left\vert E_{c}-E_{F}\right\vert /T)=\rho
_{0}^{A}\exp (\bigtriangleup /T).  \label{Eq.Arrhenius-type}
\end{equation}

Previous studies on these OSMPs reported that, due to the characteristic
arrangement of the energy levels of the involved Fe-$3d$-orbitals as well as
due to entropy arguments, a lowering of temperature often leads to a
temperature-induced crossover (OSMT) from an OSMP into a metallic phase at $%
T_{X}^{HT}$.\cite{Yi13-OSMP-AxFe2-ySe2,Yu13-OSMP-KxFe2-yK2}

In addition to the activated high-temperature process, the low temperature
resistivities of various chalcogenides\cite%
{Liu10-Fe(TeSe)-PhaseDiagram,Liu09-FeExcess-FeTeSe,Chang12-FeTe1-xSex-WeakLocalization}
are reported to exhibit another process which, in most cases, can be
approximated as a log-in-$T$ contribution: 
\begin{equation}
\rho (T)=\rho _{o}^{LT}[1+S\ln (T_{o}/T)],  \label{Eq.WeakLocalization}
\end{equation}%
the logarithmic slope $S$ is a measure of the intensity of the process while 
$T_{o}$ and $\rho _{o}^{LT}$ are characteristic, here
experimentally-determined, parameters. In contrast to the activated process,
the origin of such a log-in-$T$ behavior is not well studied; as such this
will be discussed below after the analysis of our results.

Depending on the relative strength of the above-mentioned two resistive
channels, $\rho (T)$ would assume a variety of functional forms: $\frac{%
\partial \rho }{\partial T}$ would be positive for a metallic character
while negative for any nonmetallic contribution.{\cite%
{Ando04-PhaseDiagram-HTC-Resistivity,Song11-FeSe-SUC-Normal-State}} In
addition, $\frac{\partial \rho }{\partial T}$ would be helpful in
identifying transition/crossover events (e.g. a metal-to-nonmetal, $MNM$,
crossover is manifested as a maximum in the resistivity: $\frac{\partial
\rho }{\partial T}_{T_{X}}$=0). In general, on cooling, two crossovers may
be observed: one from the high-$T$ activated regime into a metal-like phase
(metal-I) at $T_{X}^{HT}$ and another from the low-$T$ log-in-$T$ regime
into a metal-like phase (metal-II) at $T_{X}^{LT}$ (see below); it is
emphasized that, for Fe$_{1+\delta }($Te$_{1-x}$Se$_{x})$ $0.1\leq x\leq 0.5$%
, both $T_{X}^{LT}$ and $T_{X}^{HT}$ events (see also Refs. {%
\onlinecite{Liu10-Fe(TeSe)-PhaseDiagram,Liu11-Fe1+dTe1-xSex,Yadav10-Feexcess-Fe(TeSe)}%
)} are not accompanied by any visible symmetry-breaking process.

\section{Results}

Samples preparation, annealing, and measurements (namely structural,
elemental, magnetization and resistivity) of FeTe$_{1-x}$Se$_{x}$ and FeTe$%
_{1-x}$S$_{x}$ series were the same as the ones reported in previous works.%
\cite{Deguchi13-FeTeSe-beverage} The above-mentioned resistivity analysis
was applied to the measured curves of these samples: this analysis can be
readily extended to other chalcogenides.

\begin{figure}[ht]
\includegraphics[
height=6.4317cm,
width=8.1012cm
]{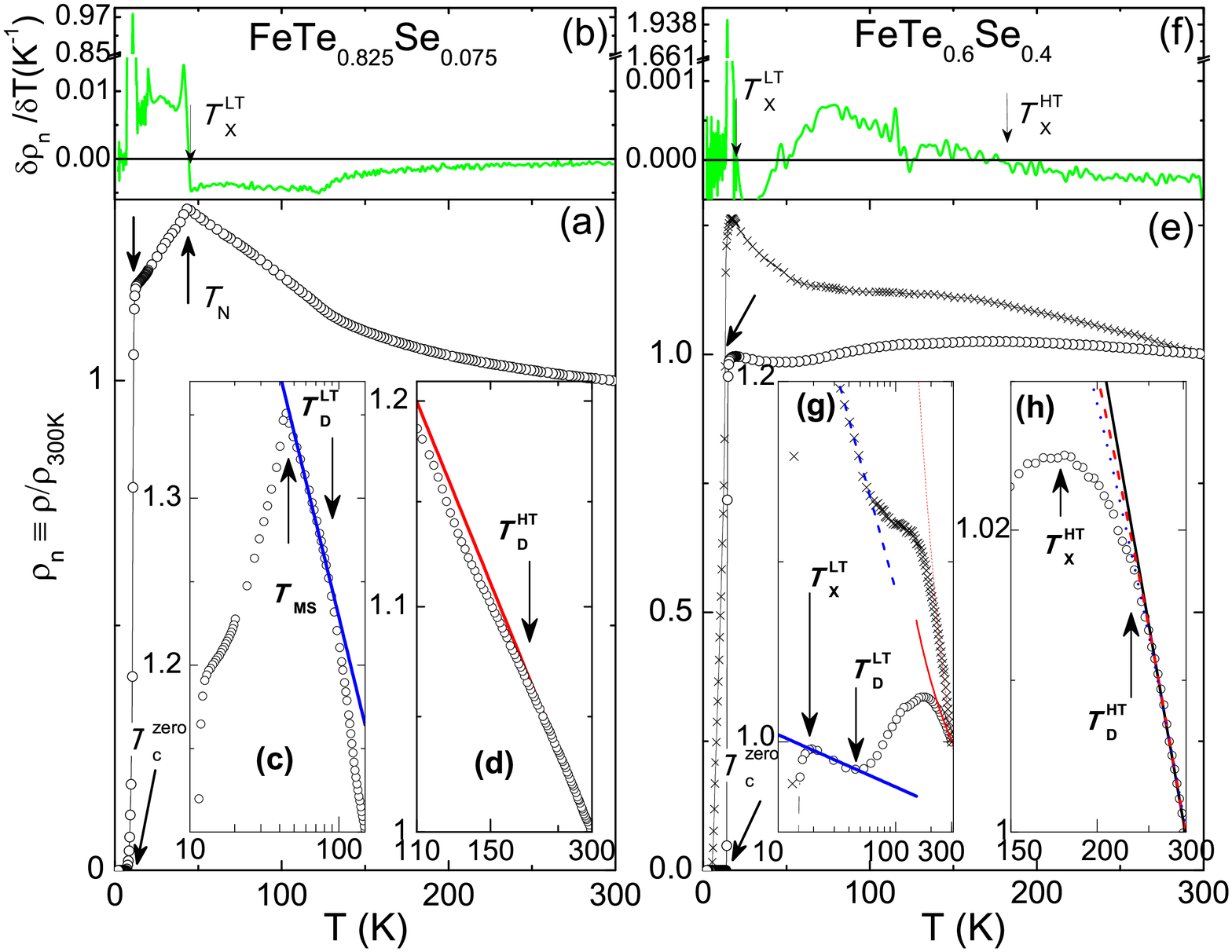}
\caption{Normalized $\protect\rho %
_{n}\equiv \protect\rho $/$\protect\rho _{300K}$ curves of FeTe$_{0.925}$Se$%
_{0.075}$ (a-d) and FeTe$_{0.6}$Se$_{0.4}$ (e-h). (a, e) $\protect\rho _{n}$ 
$vs$ $T$ curves. (b, f) $\frac{\partial \protect\rho _{n}}{\partial T}$ $vs$ 
$T$ curves showing the crossover at $T_{X}^{LT}$ and $T_{X}^{HT}$ whereat $%
\frac{\partial \protect\rho _{n}}{\partial T}$=0. (c,g) $\protect\rho _{n}$ $%
vs$ $T$ curves in a linear-log plot. The solid lines is a fit to Eq. \protect
\ref{Eq.WeakLocalization}. $S(x)$ is shown in Fig. \protect\ref%
{FeSeTe-O-anneal-PhaseDiagram}(c). (d,h) $\protect\rho _{n}$ $vs$ $T$ curves
in a log-reciprocal plot. The solid, dashed and dotted lines are, resp., a
fit\ to Eq. \protect\ref{Eq.Arrhenius-type}, Eq. \protect\ref{Eq.Mott-VRH} ($%
d$=2) and Eq.\protect\ref{Eq.Mott-VRH} ($d$=3). Notice that $T_{D}^{HT}$ is
the lower point at which $\protect\rho _{n}(T)$ starts to deviate away from
Eq. \protect\ref{Eq.Arrhenius-type} while $T_{D}^{LT}$ is the upper
deviation point from Eq. \protect\ref{Eq.WeakLocalization}. The cross
symbols in (e) and (g) represent the resistivity of the as-prepared sample:
annealing in O$_{2}$ reduces the two resistive contributions (also evident
in e.g. Ref.{\ \onlinecite{Sun2014-O-FeTeSe}}) [here $S_{\text{as-prep}}$%
=0.089(3) while $S_{\text{O}_{2}}$=0.013(1)].}
\label{Fe(TeSe)-O2-RvsT}
\end{figure}

Figure \ref{Fe(TeSe)-O2-RvsT} show $\rho (T)$ curves of two,
representatives, oxygen-annealed FeTe$_{1-x}$Se$_{x}$ samples. A closer look
at Fig. \ref{Fe(TeSe)-O2-RvsT}(b) reveals that $\frac{\partial \rho _{n}}{%
\partial T}(45<T<300K)$ exhibits, approximately, two negative values; each
is taken to indicate a distinct resistive channel, stemming from a distinct
origin, operating within a distinct temperature region, and has a distinct
thermal evolution: the one operating within 150$<T<$300K is an activated
process (see above) while the other, operating below 100K, is a log-in-$T$
process.

\begin{figure}[h]
\includegraphics[
height=6.4317cm,
width=8.1012cm
]{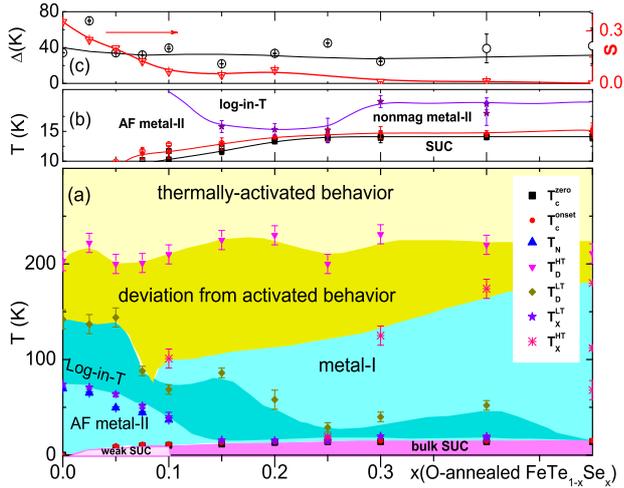}
\caption{(a) A phase diagram of FeTe$%
_{1-x}$Se$_{x}$ (0$\leq x\leq 0.5$, annealed in O$_{2}$). (b) The $x$%
-dependence of the closely-spaced $T_{c}^{zero}$, $T_{c}^{onset}$ and $%
T_{X}^{LT}$ events: in spite of their closeness, the metallic state is
(re)established well before the onset of superconductivity. (c) \textit{Left
ordinate}: $\bigtriangleup $ versus $x$ (the determination of $%
\bigtriangleup $ and $T_{VRH}$ is strongly influenced by the choice among
Eqs.\protect\ref{Eq.Mott-VRH} or \protect\ref{Eq.Arrhenius-type} and among
the limits of the available $T$-range). \textit{Right ordinate}: the
logarithmic slope $S(x)$ as obtained from the fit of Eq. \protect\ref%
{Eq.WeakLocalization}.}
\label{FeSeTe-O-anneal-PhaseDiagram}
\end{figure}

The thermal evolution of the activated process is exhibited in\ Figs. \ref%
{Fe(TeSe)-O2-RvsT}(d, h). In spite of the polycrystalline form and the
limited temperature range available for this behavior, a fit $\rho $(250$<T<$%
300K) to Eq. \ref{Eq.Mott-VRH} [see Fig. \ref{Fe(TeSe)-O2-RvsT}(h)]
indicates a simple Arrhenius expression of Eq. \ref{Eq.Arrhenius-type}
wherein the effective single parameter $\bigtriangleup $ is taken to
represent a mean localization energy separating the Fermi level from the
closest mobility edge.{\cite{Basko06-MIT}} For the particular case of FeTe$%
_{1-x}$Se$_{x}$, this $\bigtriangleup \sim $40 K is, roughly, the same for
all $x$ [see Fig. \ref{FeSeTe-O-anneal-PhaseDiagram}(c)]: such a $%
\bigtriangleup $ should not be confused with that of an activated
semiconductivity.

On the other hand, Figs. \ref{Fe(TeSe)-O2-RvsT}(c, g) indicates that $\rho
(T_{X}^{LT}<T<T_{D}^{LT})$ follows Eq. \ref{Eq.WeakLocalization};\cite%
{Liu09-FeExcess-FeTeSe,Pallecchi11-NormalState-Fe11-Fe111} wherein $S$, the
only fit parameter, is shown in Fig. \ref{FeSeTe-O-anneal-PhaseDiagram} (c).

In addition to the manifestation of two resistive channels, $\rho (T,x)$ of
FeTe$_{1-x}$Se$_{x}$ show some other finer details: (i) $\rho (T,x<0.1)$
curves are different from the ones with 0.1$\leq x\leq $0.5: the dividing
line, $x\sim $0.1, coincides with the concentration beyond which the
magnetism is suppressed.\cite%
{Mizuguchi10-FeT-Review,Deguchi12-FeX-Review,Liu10-Fe(TeSe)-PhaseDiagram}
(ii) $\rho (T,x<0.1)$ starts to deviate away from Eq. \ref{Eq.Arrhenius-type}
at $T_{D}^{HT}$; on further cooling, $\rho (T,x<0.1)$ exhibits a sharp drop
at $T_{MS}(x)$ (related to the reported magnetic and structural transition%
\cite{Martinelli10-Fe(TeSe)-PhaseDiagram}) followed by a metallic behavior.
On cooling well below $T_{MS}$, $\rho (T<T_{MS},x<0.1)$ exhibits the
transitions associated with weak and bulk superconductivity.\cite%
{Mizuguchi10-FeT-Review,Deguchi12-FeX-Review} (iii) $\rho (T,0.1\leq x\leq
0.5)$ exhibits also a deviation from Eq. \ref{Eq.Arrhenius-type} at $%
T_{D}^{HT}$, followed by a crossover into a metallic state at $T_{X}^{HT}(x)$
[see Fig. \ref{Fe(TeSe)-O2-RvsT}(f)]. On further cooling, the resistivity
once more exhibits the log-in-$T$ behavior, a crossover into a metallic
state at $T_{X}^{LT}(x)$, and finally, the bulk superconductivity at $%
T_{c}^{zero}$.

\begin{figure}[ht]
\includegraphics[
height=6.4317cm,
width=8.1012cm
]{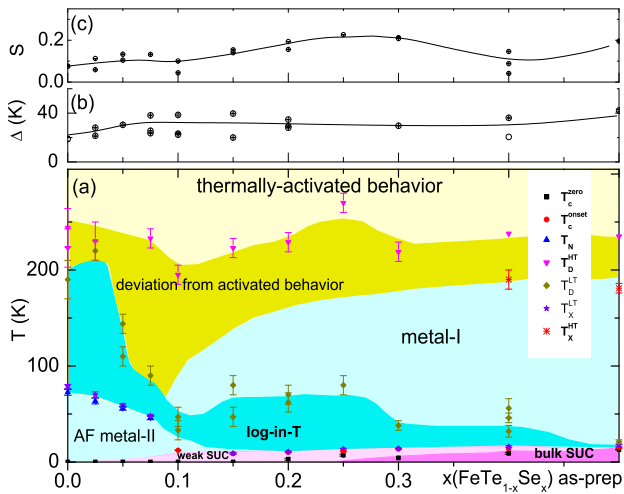}
\caption{(a) The phase diagram of
as-prepared FeTe$_{1-x}$Se$_{x}$ ($0\leq x\leq 0.5$). The difference between
this diagram and that of oxygen-annealed one (Fig. \protect\ref%
{FeSeTe-O-anneal-PhaseDiagram}) emphasizes the role of annealing and oxygen
intercalation in shaping the normal and superconducting states of these
chalcogenides. (b) Effective $\bigtriangleup $ versus $x$ as obtained from
the fit of Eq.\protect\ref{Eq.Arrhenius-type} within 250 $\leq T\leq $ 300
K. (c) $S$ versus $x$ as obtained from the fit of Eq. \protect\ref%
{Eq.WeakLocalization}.}
\label{FeSeTe-asprep-PhaseDiagram}
\end{figure}

\begin{figure}[ht]
\includegraphics[
height=6.4317cm,
width=8.1012cm
]{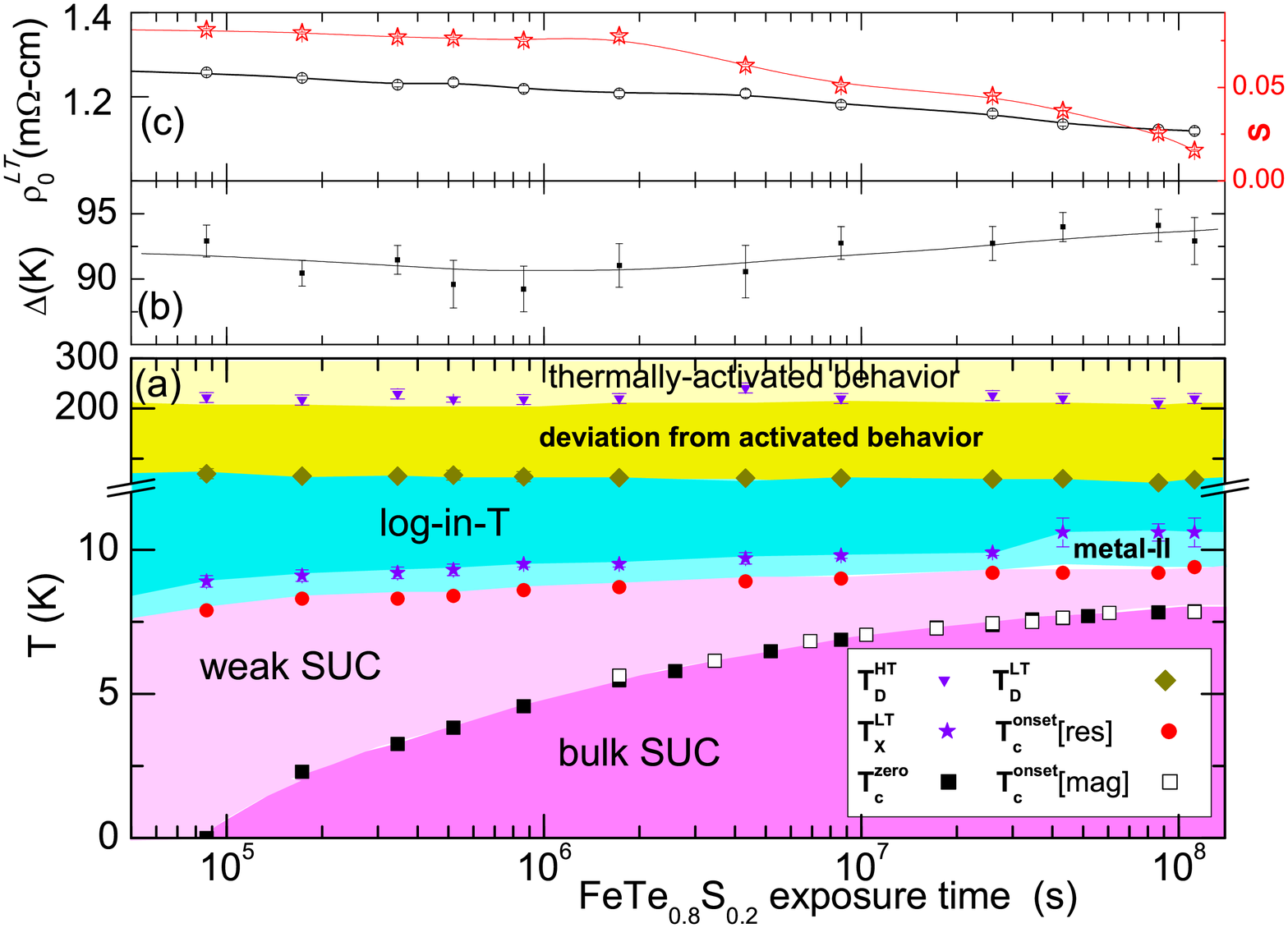}
\caption{{(a) The time evolution of the
superconducting, magnetic and transport properties of as-prepared FeTe$_{0.8}
$S$_{0.2}$. The resistivities and magnetizations (see Fig. 11 in Ref. {\ 
\onlinecite{Deguchi12-FeX-Review}}) were measured at the plotted time
intervals while the sample was being continuously exposed to air at ambient
temperature. Similar phase diagrams can be obtained for other intercalation
processes. (b) Effective\ $\bigtriangleup $ $versus$ $x$ as obtained from
the fit of Eq.\protect\ref{Eq.Arrhenius-type} within 250 $\leq T\leq $ 300
K. (c) the evolution of $\protect\rho _{o}^{LT}$ (left ordinate) and $S$
(right ordinate) versus $x$ as obtained from the fit of Eq. \protect\ref%
{Eq.WeakLocalization}.}}
\label{FeTe0.8S0.2-Phasediagram-exposure}
\end{figure}

All the above-mentioned resistivity events of oxygen-annealed FeTe$_{1-x}$Se$%
_{x}$ samples are collected in Fig. \ref{FeSeTe-O-anneal-PhaseDiagram}: in
addition to the transitions at $T_{MS}$ and $T_{c}^{zero}$ and the
crossovers at $T_{X}^{HT}$ and $T_{X}^{LT}$, we also include $T_{D}^{HT}$
and $T_{D}^{LT}$. \ Similar analyses were carried out on the resistivities
of as-prepared FeTe$_{1-x}$Se$_{x}$ samples as well as those of FeTe$_{0.8}$S%
$_{0.2}$ (representative of FeTe$_{1-x}$S$_{x}$): the obtained phase
diagrams are shown, respectively, in Figs. \ref{FeSeTe-asprep-PhaseDiagram}
and \ref{FeTe0.8S0.2-Phasediagram-exposure}.

The pressure influence on the superconducting and normal-state properties of 
these chalcogenides was extensively studied [see e.g., Fe$_{1+\delta }$Te,%
\cite{Okada09-PhaseTrans-FeTe} Fe$_{1+\delta }$Se,\cite%
{Mizuguchi08-FeSe-27K-pressure,Margadonna09-FeSe-Pressure,Imai09-FeSe-HighPressure,Medvedev09-FeSe-Pressure,Li09-FeSex-Pressure, Braithwaite09-FeSe-HighPressure,Garbarino09-FeSe-HighPressure}
Fe$_{1+\delta }$S,\cite{Denholme14-FeS-metallicity} and Fe$_{1+\delta }($Te$%
_{1-x}$Se$_{x})$.\cite%
{Horigane09-FeTe05Se05,Tsoi09-FeSe05Te05-HighPressure,Huang09-FeSe088-FeSe05Te05-Pressure,Stemshorn10-FeSe-HighPressure}%
] In general, $\rho (T_{c}<T<$300K$)$ is strongly reduced by $P$. It is
noticed that, even for pressures leading to amorphization,\cite%
{Stemshorn10-FeSe-HighPressure} the evolution of $\rho (T_{c}<T<$300K$,P)$
does not support an interpretation in terms of a closure of a semiconducting
gap. Rather, such an influence can be envisaged as a reduction of the two
resistive channels and an enhancement of the metallic state: For Fe$%
_{1+\delta }$Te, the metallic state (stabilized below $T_{MS}$ at ambient
pressure) is enhanced to $\sim $240 K at 7 GPa,\cite{Okada09-PhaseTrans-FeTe}
while for Fe$_{1+\delta }$S, the metallic state (stabilized above 300 K at
ambient pressure) is extended downwards to $\sim $70 K at $\sim $3 GPa.\cite%
{Denholme14-FeS-metallicity}

Finally, $T_{N,mag}$ and $T_{c,mag}^{onset}$, as determined from
magnetization curves (not shown), are shown in the respective phase
diagrams; evidently $T_{c,mag}^{onset}$ equals $T_{c,\rho }^{zero}$; as such 
$T_{c,\rho }^{onset}$ should not be taken as an onset of superconductivity,
rather it is $\sim T_{X}^{LT}$ (see below).

\section{Discussion and conclusions}

We showed a consistent and a unified analysis of $\rho (x,P,T<$300K$)$ based
on consideration of an activated and a log-in-$T$ processes. Various $x$%
-dependent transition/crossover events ($T_{MS}$, $T_{N}$, $T_{c}^{zero}$, $%
T_{X}^{HT}$, $T_{X}^{LT}$, $T_{D}^{HT}$ and $T_{D}^{LT}$) were identified
and these points were used to construct the corresponding $x$-$T$ phase
diagrams; that these diagrams are topologically equivalent to the earlier
reported ones\cite%
{Mizuguchi10-FeT-Review,Deguchi12-FeX-Review,Katayama10-Fe(SeTe)-phaseDiagram,Liu10-Fe(TeSe)-PhaseDiagram,Sales09-FeTeSe-SUC,*Tropeano10-Fe(TeSe)-transport,Fang08-Fe(SeTe),Dong11-PhaseDiagram-FeTeSe,Kawasaki12-Fe(TeSe)-O-anneal-PhaseDiagram,Martinelli10-Fe(TeSe)-PhaseDiagram}
emphasizes the role of these channels in shaping the phase diagrams.

The origin of the high-$T$ activated process, as already discussed, is
related to the presence of an orbital-selective Mott phase.\cite%
{Yi13-OSMP-AxFe2-ySe2,Yu13-OSMP-KxFe2-yK2,Craco11-Mott-KxFe2-ySe2,Yin12-powerLaw-Fe-Chalcogenides,Bascones12-OSMP-Magnetic-Fe-SUCs}
A careful analysis of this process in Fe$_{1+\delta }X$ suggests that Fe$%
_{1+\delta }$S is at the vicinity of Mott transition, Fe$_{1+\delta }$Se is
at a metallic side, while Fe$_{1+\delta }$Te and Fe$_{1+\delta }$Te$%
_{1-x}Y_{x}$ solid solutions are at an intermediate orbital-selective Mott
phase. A variation in any of the control parameters (crystal structure,
intercalation, pressure, defects, or filling) would turn the normal-state
into either a metal-like, an activated OSMP, or a Mott insulator.\cite%
{Yi13-OSMP-AxFe2-ySe2,Yu13-OSMP-KxFe2-yK2,Wang14-MIT-RbFe2Se2,Craco11-Mott-KxFe2-ySe2,Yin12-powerLaw-Fe-Chalcogenides,Bascones12-OSMP-Magnetic-Fe-SUCs}
This was elegantly illustrated in the case of $A$Fe$_{2-y}$Se$_{2}$.\cite%
{Fang11-(TlK)FexSe2,Yi13-OSMP-AxFe2-ySe2}

The origin of the low-$T$ localization, on the other hand, can be deduced
from the scanning tunneling microscopy studies of Machida \textit{et al.}%
\cite{Machida13-FeTe-STM} which revealed (at 4.2 K, within the metallic
state of Fe$_{1+\delta }$Te) the presence of an inhomogeneous\ distribution
of local density of states (LDOS) around the randomly-distributed defect
centres (suggested to be excess Fe): LDOS is increased around these centres
while depleted away from them. Scattering from such nonperiodic potentials
is assumed to be the cause of the low-$T$ process.\cite%
{Liu10-Fe(TeSe)-PhaseDiagram,Liu09-FeExcess-FeTeSe,Machida13-FeTe-excess-Fe,Chang12-FeTe1-xSex-WeakLocalization,Hu13-Fe1+y(Te1-xSex)}
Indeed, the strength of such a process is strongly controlled by the
concentration of defects: this is well illustrated in the correlation of $%
\rho (T<T_{MS})$ with $\delta $ in Fe$_{1+\delta }$Te [{%
\onlinecite{Machida13-FeTe-excess-Fe}]} as well as in the correlation of $S$(%
$x)$ (see Fig. \ref{FeSeTe-O-anneal-PhaseDiagram}) with the excess Fe or
chalcogen deficiency.\cite{Sun2014-O-FeTeSe} In general, the control of
scattering centres can be effected by annealing (Fig. \ref%
{FeSeTe-O-anneal-PhaseDiagram} and Ref. {\onlinecite{Cao11-FeExcess-FeTeSe}}%
), applying pressure (Refs.{%
\onlinecite{Okada09-PhaseTrans-FeTe,
Mizuguchi08-FeSe-27K-pressure,
Margadonna09-FeSe-Pressure,Imai09-FeSe-HighPressure,Medvedev09-FeSe-Pressure,Li09-FeSex-Pressure, Braithwaite09-FeSe-HighPressure,Garbarino09-FeSe-HighPressure,Horigane09-FeTe05Se05, Tsoi09-FeSe05Te05-HighPressure, Huang09-FeSe088-FeSe05Te05-Pressure, Stemshorn10-FeSe-HighPressure, Denholme14-FeS-metallicity}%
}), or intercalation (Fig. \ref{FeTe0.8S0.2-Phasediagram-exposure} and Ref. {%
\onlinecite{Deguchi12-FeX-Review}}).

This defect-driven low-$T$ process is manifested here as, though not limited
to,\ a log-in-$T$ term; three possible mechanisms can be suggested for this
log-in-$T$ behavior: (i) a typical Kondo or any of the formally analogous
processes arising from scattering against nonmagnetic degenerate-state
impurities,\cite%
{Taraphder91-HF-negative-U-Anderson-Model,Ralph92-Kondo-No-MagImpurity} (ii)
Anderson weak localization wherein defects lead to a coherent backscattering
of non-interacting electrons,\cite%
{Anderson79-NonlinearConductance,Dolan79-nonlinearConductance} or (iii)
Altshuller-Aronov\cite{Altshuler83-Localization-e-e} weak localization
wherein the system of interacting electrons are subjected to a random
potential. The following considerations elect the Altshuller-Aronov weak
localization process: First, these chalcogenides are structurally formed
from Fe$_{2}X_{2}$ layers that are coupled to their neighbors by weak van
der Waals forces. Second, the observation of stronger correlation effects in
chalcogenides\cite{Singh12-Chalcogenides-Review} suggests a scenario of 
\textit{interacting} electrons in a random potential which may arise from a
randomly-distributed (paramagnetic and/or nonmagnetic) impurities.

The identification of the origins of these two processes makes it easy to
trace down the influence of perturbations (such as pressure, intercalation,
annealing, ..) on the boundary lines at $T_{X}^{LT}$ and $T_{X}^{HT}$ of the
studied phase diagrams. While the origin of the crossover at $T_{X}^{HT}$ is
already discussed, that of $T_{X}^{LT}$ is most probably related to the
fluctuation-induced formation of incoherent Cooper pairs which is favored by
disorder and lower dimensionality of chalcogenides:\cite%
{Sacepe10-PseudoGap-TiN,Lee06-Doping-Mott-Insulator-Review} due to the onset
of these fluctuations, the metallic-like character is (re)established well
before the onset of superconductivity (see Figs. \ref%
{FeSeTe-O-anneal-PhaseDiagram}, \ref{FeSeTe-asprep-PhaseDiagram}, \ref%
{FeTe0.8S0.2-Phasediagram-exposure}). This feature is reminiscent of the
situation in both nonconventional underdoped $HT_{c}$, see e.g. Ref. {%
\onlinecite{Ando04-PhaseDiagram-HTC-Resistivity}}, and conventional $LT_{c}$
superconductors.\cite{Sacepe10-PseudoGap-TiN} Accordingly, $T_{X}^{LT}$ may
be identified as the onset point, $T^{\ast }$, of a pseudogap phase.\cite%
{Pallecchi11-NormalState-Fe11-Fe111} It is emphasized that this
low-temperature disorder-driven metal-nonmetal transition is different from
the high-temperature Mott metal-nonmetal transition.

The competition between superconductivity and localization, observed above,
may be discussed in terms of a competition between an attractive ($V_{p}$)
and a repulsive ($V_{c}$) interactions. On assuming an effective coupling $%
\lambda _{eff}=\lambda -\mu ^{\ast }$ wherein $\lambda =V_{p}N_{E_{F}}$ and $%
\mu ^{\ast }=\frac{V_{c}N_{E_{F}}}{1+V_{c}N_{E_{F}}\ln (\frac{\omega _{_{c}}%
}{\omega _{D}})}$ (all terms have their usual meaning),\cite%
{Ketterson-Song-SUC-textbook} then (i) an increase in, say, disorder would
augment $\mu ^{\ast }$ which, being in competition with superconductivity,
would degrade and, above a critical value, eventually quench the
superconductivity.\cite{Sacepe10-PseudoGap-TiN} A further increase of
disorder would eventually lead to a stronger Anderson-type localization as
best illustrated in Fe$_{1.01-x}$Cu$_{x}$Se.\cite{Williams09-(FeCu)Se-MIT}
(ii) the manifestation of $T_{X}^{LT}$ and the associated metal-II phase is
conditioned by the strength of the attractive term: if no or weaker
superconductivity, then no $T_{X}^{LT}$ event.

Finally, based on the above-mentioned remarks, it is easy to understand (i)
why the resistivity behavior of most Fe-based chalcogenides are neither
truly metallic nor truly insulating, (ii) why some events are not exhibited
in the reported resistivities of the as-prepared or annealed/intercalated
samples, and (iii) why the reported superconducting and magnetic phase
diagrams are strongly influenced by sample conditions (compare Figs. \ref%
{FeSeTe-O-anneal-PhaseDiagram} and \ref{FeSeTe-asprep-PhaseDiagram}).

\begin{acknowledgments}
We acknowledges fruitful discussions with D Edwards, N Trivedi, J
Albuquerque, C M Chaves, A Troper, M A C Gusmao, L Craco, C Lewenkopf, and S
Das Sarma.
\end{acknowledgments}

\bibliographystyle{apsrev4-1}
\bibliography{FeS-mackinawite,intermetallic,InterplaySupMag,Localization,pnictides,PseudoGap,spinfluctuation,SupClassic}

\begin{thebibliography}{64}%
\makeatletter
\providecommand \@ifxundefined [1]{%
 \@ifx{#1\undefined}
}%
\providecommand \@ifnum [1]{%
 \ifnum #1\expandafter \@firstoftwo
 \else \expandafter \@secondoftwo
 \fi
}%
\providecommand \@ifx [1]{%
 \ifx #1\expandafter \@firstoftwo
 \else \expandafter \@secondoftwo
 \fi
}%
\providecommand \natexlab [1]{#1}%
\providecommand \enquote  [1]{``#1''}%
\providecommand \bibnamefont  [1]{#1}%
\providecommand \bibfnamefont [1]{#1}%
\providecommand \citenamefont [1]{#1}%
\providecommand \href@noop [0]{\@secondoftwo}%
\providecommand \href [0]{\begingroup \@sanitize@url \@href}%
\providecommand \@href[1]{\@@startlink{#1}\@@href}%
\providecommand \@@href[1]{\endgroup#1\@@endlink}%
\providecommand \@sanitize@url [0]{\catcode `\\12\catcode `\$12\catcode
  `\&12\catcode `\#12\catcode `\^12\catcode `\_12\catcode `\%12\relax}%
\providecommand \@@startlink[1]{}%
\providecommand \@@endlink[0]{}%
\providecommand \url  [0]{\begingroup\@sanitize@url \@url }%
\providecommand \@url [1]{\endgroup\@href {#1}{\urlprefix }}%
\providecommand \urlprefix  [0]{URL }%
\providecommand \Eprint [0]{\href }%
\providecommand \doibase [0]{http://dx.doi.org/}%
\providecommand \selectlanguage [0]{\@gobble}%
\providecommand \bibinfo  [0]{\@secondoftwo}%
\providecommand \bibfield  [0]{\@secondoftwo}%
\providecommand \translation [1]{[#1]}%
\providecommand \BibitemOpen [0]{}%
\providecommand \bibitemStop [0]{}%
\providecommand \bibitemNoStop [0]{.\EOS\space}%
\providecommand \EOS [0]{\spacefactor3000\relax}%
\providecommand \BibitemShut  [1]{\csname bibitem#1\endcsname}%
\let\auto@bib@innerbib\@empty
\bibitem [{\citenamefont {Mizuguchi}\ and\ \citenamefont
  {Takano}(2010)}]{Mizuguchi10-FeT-Review}%
  \BibitemOpen
  \bibfield  {author} {\bibinfo {author} {\bibfnamefont {Y.}~\bibnamefont
  {Mizuguchi}}\ and\ \bibinfo {author} {\bibfnamefont {Y.}~\bibnamefont
  {Takano}},\ }\href@noop {} {\bibfield  {journal} {\bibinfo  {journal} {J.
  Phys. Soc. Jpn.}\ }\textbf {\bibinfo {volume} {79}},\ \bibinfo {pages}
  {102001} (\bibinfo {year} {2010})}\BibitemShut {NoStop}%
\bibitem [{\citenamefont {Deguchi}\ \emph {et~al.}(2012)\citenamefont
  {Deguchi}, \citenamefont {Takano},\ and\ \citenamefont
  {Mizuguchi}}]{Deguchi12-FeX-Review}%
  \BibitemOpen
  \bibfield  {author} {\bibinfo {author} {\bibfnamefont {K.}~\bibnamefont
  {Deguchi}}, \bibinfo {author} {\bibfnamefont {Y.}~\bibnamefont {Takano}}, \
  and\ \bibinfo {author} {\bibfnamefont {Y.}~\bibnamefont {Mizuguchi}},\
  }\href@noop {} {\bibfield  {journal} {\bibinfo  {journal} {Sci. Technol. Adv.
  Mater.}\ }\textbf {\bibinfo {volume} {13}},\ \bibinfo {pages} {054303}
  (\bibinfo {year} {2012})}\BibitemShut {NoStop}%
\bibitem [{\citenamefont {Liu}\ \emph {et~al.}(2010)\citenamefont {Liu} \emph
  {et~al.}}]{Liu10-Fe(TeSe)-PhaseDiagram}%
  \BibitemOpen
  \bibfield  {author} {\bibinfo {author} {\bibfnamefont {T.~J.}\ \bibnamefont
  {Liu}} \emph {et~al.},\ }\href@noop {} {\bibfield  {journal} {\bibinfo
  {journal} {Nat. Mater.}\ }\textbf {\bibinfo {volume} {9}},\ \bibinfo {pages}
  {716} (\bibinfo {year} {2010})}\BibitemShut {NoStop}%
\bibitem [{\citenamefont {Katayama}\ \emph {et~al.}(2010)\citenamefont
  {Katayama}, \citenamefont {Ji}, \citenamefont {Louca}, \citenamefont {Lee},
  \citenamefont {Fujita}, \citenamefont {Sato}, \citenamefont {Wen},
  \citenamefont {Xu}, \citenamefont {Gu}, \citenamefont {Xu}, \citenamefont
  {Lin}, \citenamefont {Enoki}, \citenamefont {Chang}, \citenamefont {Yamada},\
  and\ \citenamefont {Tranquada}}]{Katayama10-Fe(SeTe)-phaseDiagram}%
  \BibitemOpen
  \bibfield  {author} {\bibinfo {author} {\bibfnamefont {N.}~\bibnamefont
  {Katayama}}, \bibinfo {author} {\bibfnamefont {S.}~\bibnamefont {Ji}},
  \bibinfo {author} {\bibfnamefont {D.}~\bibnamefont {Louca}}, \bibinfo
  {author} {\bibfnamefont {S.}~\bibnamefont {Lee}}, \bibinfo {author}
  {\bibfnamefont {M.}~\bibnamefont {Fujita}}, \bibinfo {author} {\bibfnamefont
  {T.~J.}\ \bibnamefont {Sato}}, \bibinfo {author} {\bibfnamefont
  {J.}~\bibnamefont {Wen}}, \bibinfo {author} {\bibfnamefont {Z.}~\bibnamefont
  {Xu}}, \bibinfo {author} {\bibfnamefont {G.}~\bibnamefont {Gu}}, \bibinfo
  {author} {\bibfnamefont {G.}~\bibnamefont {Xu}}, \bibinfo {author}
  {\bibfnamefont {Z.}~\bibnamefont {Lin}}, \bibinfo {author} {\bibfnamefont
  {M.}~\bibnamefont {Enoki}}, \bibinfo {author} {\bibfnamefont
  {S.}~\bibnamefont {Chang}}, \bibinfo {author} {\bibfnamefont
  {K.}~\bibnamefont {Yamada}}, \ and\ \bibinfo {author} {\bibfnamefont {J.~M.}\
  \bibnamefont {Tranquada}},\ }\href@noop {} {\bibfield  {journal} {\bibinfo
  {journal} {J. Phys. Soc. Jpn.}\ }\textbf {\bibinfo {volume} {79}},\ \bibinfo
  {pages} {113702} (\bibinfo {year} {2010})}\BibitemShut {NoStop}%
\bibitem [{\citenamefont {Sales}\ \emph {et~al.}(2009)\citenamefont {Sales},
  \citenamefont {Sefat}, \citenamefont {McGuire}, \citenamefont {Jin},
  \citenamefont {Mandrus},\ and\ \citenamefont
  {Mozharivskyj}}]{Sales09-FeTeSe-SUC}%
  \BibitemOpen
  \bibfield  {author} {\bibinfo {author} {\bibfnamefont {B.~C.}\ \bibnamefont
  {Sales}}, \bibinfo {author} {\bibfnamefont {A.~S.}\ \bibnamefont {Sefat}},
  \bibinfo {author} {\bibfnamefont {M.~A.}\ \bibnamefont {McGuire}}, \bibinfo
  {author} {\bibfnamefont {R.~Y.}\ \bibnamefont {Jin}}, \bibinfo {author}
  {\bibfnamefont {D.}~\bibnamefont {Mandrus}}, \ and\ \bibinfo {author}
  {\bibfnamefont {Y.}~\bibnamefont {Mozharivskyj}},\ }\href@noop {} {\bibfield
  {journal} {\bibinfo  {journal} {Phys. Rev. B}\ }\textbf {\bibinfo {volume}
  {79}},\ \bibinfo {pages} {094521} (\bibinfo {year} {2009})}\BibitemShut
  {NoStop}%
\bibitem [{\citenamefont {Tropeano}\ \emph {et~al.}(2010)\citenamefont
  {Tropeano}, \citenamefont {Pallecchi}, \citenamefont {Cimberle},
  \citenamefont {Ferdeghini}, \citenamefont {Lamura}, \citenamefont {Vignolo},
  \citenamefont {Martinelli}, \citenamefont {Palenzona},\ and\ \citenamefont
  {Putti}}]{Tropeano10-Fe(TeSe)-transport}%
  \BibitemOpen
  \bibfield  {author} {\bibinfo {author} {\bibfnamefont {M.}~\bibnamefont
  {Tropeano}}, \bibinfo {author} {\bibfnamefont {I.}~\bibnamefont {Pallecchi}},
  \bibinfo {author} {\bibfnamefont {M.~R.}\ \bibnamefont {Cimberle}}, \bibinfo
  {author} {\bibfnamefont {C.}~\bibnamefont {Ferdeghini}}, \bibinfo {author}
  {\bibfnamefont {G.}~\bibnamefont {Lamura}}, \bibinfo {author} {\bibfnamefont
  {M.}~\bibnamefont {Vignolo}}, \bibinfo {author} {\bibfnamefont
  {A.}~\bibnamefont {Martinelli}}, \bibinfo {author} {\bibfnamefont
  {A.}~\bibnamefont {Palenzona}}, \ and\ \bibinfo {author} {\bibfnamefont
  {M.}~\bibnamefont {Putti}},\ }\href@noop {} {\bibfield  {journal} {\bibinfo
  {journal} {Supercond. Sci. Technol.}\ }\textbf {\bibinfo {volume} {23}},\
  \bibinfo {pages} {054001} (\bibinfo {year} {2010})}\BibitemShut {NoStop}%
\bibitem [{\citenamefont {Guo}\ \emph {et~al.}(2010)\citenamefont {Guo},
  \citenamefont {Jin}, \citenamefont {Wang}, \citenamefont {Wang},
  \citenamefont {Zhu}, \citenamefont {Zhou}, \citenamefont {He},\ and\
  \citenamefont {Chen}}]{Guo10-KxFe2Se2-Sup}%
  \BibitemOpen
  \bibfield  {author} {\bibinfo {author} {\bibfnamefont {J.}~\bibnamefont
  {Guo}}, \bibinfo {author} {\bibfnamefont {S.}~\bibnamefont {Jin}}, \bibinfo
  {author} {\bibfnamefont {G.}~\bibnamefont {Wang}}, \bibinfo {author}
  {\bibfnamefont {S.}~\bibnamefont {Wang}}, \bibinfo {author} {\bibfnamefont
  {K.}~\bibnamefont {Zhu}}, \bibinfo {author} {\bibfnamefont {T.}~\bibnamefont
  {Zhou}}, \bibinfo {author} {\bibfnamefont {M.}~\bibnamefont {He}}, \ and\
  \bibinfo {author} {\bibfnamefont {X.}~\bibnamefont {Chen}},\ }\href@noop {}
  {\bibfield  {journal} {\bibinfo  {journal} {Phys. Rev. B}\ }\textbf {\bibinfo
  {volume} {82}},\ \bibinfo {pages} {180520} (\bibinfo {year}
  {2010})}\BibitemShut {NoStop}%
\bibitem [{\citenamefont {Fang}\ \emph {et~al.}(2011)\citenamefont {Fang},
  \citenamefont {Wang}, \citenamefont {Dong}, \citenamefont {Li}, \citenamefont
  {Feng}, \citenamefont {Chen},\ and\ \citenamefont
  {Yuan}}]{Fang11-(TlK)FexSe2}%
  \BibitemOpen
  \bibfield  {author} {\bibinfo {author} {\bibfnamefont {M.-H.}\ \bibnamefont
  {Fang}}, \bibinfo {author} {\bibfnamefont {H.-D.}\ \bibnamefont {Wang}},
  \bibinfo {author} {\bibfnamefont {C.-H.}\ \bibnamefont {Dong}}, \bibinfo
  {author} {\bibfnamefont {Z.-J.}\ \bibnamefont {Li}}, \bibinfo {author}
  {\bibfnamefont {C.-M.}\ \bibnamefont {Feng}}, \bibinfo {author}
  {\bibfnamefont {J.}~\bibnamefont {Chen}}, \ and\ \bibinfo {author}
  {\bibfnamefont {H.~Q.}\ \bibnamefont {Yuan}},\ }\href@noop {} {\bibfield
  {journal} {\bibinfo  {journal} {Europhys. Lett.}\ }\textbf {\bibinfo {volume}
  {94}},\ \bibinfo {pages} {27009} (\bibinfo {year} {2011})}\BibitemShut
  {NoStop}%
\bibitem [{\citenamefont {Ying}\ \emph {et~al.}(2012)\citenamefont {Ying},
  \citenamefont {Chen}, \citenamefont {Wang}, \citenamefont {Jin},
  \citenamefont {Zhou}, \citenamefont {Lai}, \citenamefont {Zhang},\ and\
  \citenamefont {Wang}}]{Ying12-AxFe2Se2}%
  \BibitemOpen
  \bibfield  {author} {\bibinfo {author} {\bibfnamefont {T.~P.}\ \bibnamefont
  {Ying}}, \bibinfo {author} {\bibfnamefont {X.~L.}\ \bibnamefont {Chen}},
  \bibinfo {author} {\bibfnamefont {G.}~\bibnamefont {Wang}}, \bibinfo {author}
  {\bibfnamefont {S.~F.}\ \bibnamefont {Jin}}, \bibinfo {author} {\bibfnamefont
  {T.~T.}\ \bibnamefont {Zhou}}, \bibinfo {author} {\bibfnamefont {X.~F.}\
  \bibnamefont {Lai}}, \bibinfo {author} {\bibfnamefont {H.}~\bibnamefont
  {Zhang}}, \ and\ \bibinfo {author} {\bibfnamefont {W.~Y.}\ \bibnamefont
  {Wang}},\ }\href@noop {} {\bibfield  {journal} {\bibinfo  {journal} {Nature
  Sci. Rep.}\ }\textbf {\bibinfo {volume} {2}},\ \bibinfo {pages} {426}
  (\bibinfo {year} {2012})}\BibitemShut {NoStop}%
\bibitem [{\citenamefont {Wang}\ \emph {et~al.}(2011)\citenamefont {Wang},
  \citenamefont {He}, \citenamefont {Xia},\ and\ \citenamefont
  {Chen}}]{Wang11-doping-KxFe2-ySe2}%
  \BibitemOpen
  \bibfield  {author} {\bibinfo {author} {\bibfnamefont {D.~M.}\ \bibnamefont
  {Wang}}, \bibinfo {author} {\bibfnamefont {J.~B.}\ \bibnamefont {He}},
  \bibinfo {author} {\bibfnamefont {T.-L.}\ \bibnamefont {Xia}}, \ and\
  \bibinfo {author} {\bibfnamefont {G.~F.}\ \bibnamefont {Chen}},\ }\href@noop
  {} {\bibfield  {journal} {\bibinfo  {journal} {Phys. Rev. B}\ }\textbf
  {\bibinfo {volume} {83}},\ \bibinfo {pages} {132502} (\bibinfo {year}
  {2011})}\BibitemShut {NoStop}%
\bibitem [{\citenamefont {Wang}\ \emph {et~al.}(2014)\citenamefont {Wang},
  \citenamefont {Schmidt}, \citenamefont {Tsurkan}, \citenamefont {Greger},
  \citenamefont {Vollhard}, \citenamefont {Loid},\ and\ \citenamefont
  {Deisenhofer}}]{Wang14-MIT-RbFe2Se2}%
  \BibitemOpen
  \bibfield  {author} {\bibinfo {author} {\bibfnamefont {Z.}~\bibnamefont
  {Wang}}, \bibinfo {author} {\bibfnamefont {M.}~\bibnamefont {Schmidt}},
  \bibinfo {author} {\bibfnamefont {J.~F.~V.}\ \bibnamefont {Tsurkan}},
  \bibinfo {author} {\bibfnamefont {M.}~\bibnamefont {Greger}}, \bibinfo
  {author} {\bibfnamefont {D.}~\bibnamefont {Vollhard}}, \bibinfo {author}
  {\bibfnamefont {A.}~\bibnamefont {Loid}}, \ and\ \bibinfo {author}
  {\bibfnamefont {J.}~\bibnamefont {Deisenhofer}},\ }\href@noop {} {\bibfield
  {journal} {\bibinfo  {journal} {Nature Commun}\ }\textbf {\bibinfo {volume}
  {5}},\ \bibinfo {pages} {3202} (\bibinfo {year} {2014})}\BibitemShut
  {NoStop}%
\bibitem [{\citenamefont {Chen}\ \emph {et~al.}(2011)\citenamefont {Chen},
  \citenamefont {Xu}, \citenamefont {Ge}, \citenamefont {Zhang}, \citenamefont
  {Ye}, \citenamefont {Yang}, \citenamefont {Jiang}, \citenamefont {Xie},
  \citenamefont {Che}, \citenamefont {Zhang}, \citenamefont {Wang},
  \citenamefont {Chen}, \citenamefont {Shen}, \citenamefont {Hu},\ and\
  \citenamefont {Feng}}]{Chen11-KxFe2-ySe2}%
  \BibitemOpen
  \bibfield  {author} {\bibinfo {author} {\bibfnamefont {F.}~\bibnamefont
  {Chen}}, \bibinfo {author} {\bibfnamefont {M.}~\bibnamefont {Xu}}, \bibinfo
  {author} {\bibfnamefont {Q.~Q.}\ \bibnamefont {Ge}}, \bibinfo {author}
  {\bibfnamefont {Y.}~\bibnamefont {Zhang}}, \bibinfo {author} {\bibfnamefont
  {Z.~R.}\ \bibnamefont {Ye}}, \bibinfo {author} {\bibfnamefont {L.~X.}\
  \bibnamefont {Yang}}, \bibinfo {author} {\bibfnamefont {J.}~\bibnamefont
  {Jiang}}, \bibinfo {author} {\bibfnamefont {B.~P.}\ \bibnamefont {Xie}},
  \bibinfo {author} {\bibfnamefont {R.~C.}\ \bibnamefont {Che}}, \bibinfo
  {author} {\bibfnamefont {M.}~\bibnamefont {Zhang}}, \bibinfo {author}
  {\bibfnamefont {A.~F.}\ \bibnamefont {Wang}}, \bibinfo {author}
  {\bibfnamefont {X.~H.}\ \bibnamefont {Chen}}, \bibinfo {author}
  {\bibfnamefont {D.~W.}\ \bibnamefont {Shen}}, \bibinfo {author}
  {\bibfnamefont {J.~P.}\ \bibnamefont {Hu}}, \ and\ \bibinfo {author}
  {\bibfnamefont {D.~L.}\ \bibnamefont {Feng}},\ }\href@noop {} {\bibfield
  {journal} {\bibinfo  {journal} {Phys. Rev. X}\ }\textbf {\bibinfo {volume}
  {1}},\ \bibinfo {pages} {021020} (\bibinfo {year} {2011})}\BibitemShut
  {NoStop}%
\bibitem [{\citenamefont {Singh}(2012)}]{Singh12-Chalcogenides-Review}%
  \BibitemOpen
  \bibfield  {author} {\bibinfo {author} {\bibfnamefont {D.~J.}\ \bibnamefont
  {Singh}},\ }\href@noop {} {\bibfield  {journal} {\bibinfo  {journal} {Sci.
  Technol. Adv. Mater}\ }\textbf {\bibinfo {volume} {13}},\ \bibinfo {pages}
  {054304} (\bibinfo {year} {2012})}\BibitemShut {NoStop}%
\bibitem [{\citenamefont {Bertaut}\ \emph {et~al.}(1965)\citenamefont
  {Bertaut}, \citenamefont {Burlet},\ and\ \citenamefont
  {Chappert}}]{Bertaut65}%
  \BibitemOpen
  \bibfield  {author} {\bibinfo {author} {\bibfnamefont {E.~F.}\ \bibnamefont
  {Bertaut}}, \bibinfo {author} {\bibfnamefont {P.}~\bibnamefont {Burlet}}, \
  and\ \bibinfo {author} {\bibfnamefont {J.}~\bibnamefont {Chappert}},\
  }\href@noop {} {\bibfield  {journal} {\bibinfo  {journal} {Solid State
  Comm.}\ }\textbf {\bibinfo {volume} {3}},\ \bibinfo {pages} {335} (\bibinfo
  {year} {1965})}\BibitemShut {NoStop}%
\bibitem [{\citenamefont {Denholme}\ \emph {et~al.}(2014)\citenamefont
  {Denholme} \emph {et~al.}}]{Denholme14-FeS-metallicity}%
  \BibitemOpen
  \bibfield  {author} {\bibinfo {author} {\bibfnamefont {S.}~\bibnamefont
  {Denholme}} \emph {et~al.},\ }\href@noop {} {\bibfield  {journal} {\bibinfo
  {journal} {Sci. Technol. Adv. Mater.}\ } (\bibinfo {year} {2014})},\ \bibinfo
  {note} {in press}\BibitemShut {NoStop}%
\bibitem [{\citenamefont {McQueen}\ \emph {et~al.}(2009)\citenamefont
  {McQueen}, \citenamefont {Williams}, \citenamefont {Stephens}, \citenamefont
  {Tao}, \citenamefont {Zhu}, \citenamefont {Ksenofontov}, \citenamefont
  {Casper}, \citenamefont {Felser},\ and\ \citenamefont
  {Cava}}]{McQueen09-FeSe-StrTransition}%
  \BibitemOpen
  \bibfield  {author} {\bibinfo {author} {\bibfnamefont {T.~M.}\ \bibnamefont
  {McQueen}}, \bibinfo {author} {\bibfnamefont {A.~J.}\ \bibnamefont
  {Williams}}, \bibinfo {author} {\bibfnamefont {P.~W.}\ \bibnamefont
  {Stephens}}, \bibinfo {author} {\bibfnamefont {J.}~\bibnamefont {Tao}},
  \bibinfo {author} {\bibfnamefont {Y.}~\bibnamefont {Zhu}}, \bibinfo {author}
  {\bibfnamefont {V.}~\bibnamefont {Ksenofontov}}, \bibinfo {author}
  {\bibfnamefont {F.}~\bibnamefont {Casper}}, \bibinfo {author} {\bibfnamefont
  {C.}~\bibnamefont {Felser}}, \ and\ \bibinfo {author} {\bibfnamefont {R.~J.}\
  \bibnamefont {Cava}},\ }\href@noop {} {\bibfield  {journal} {\bibinfo
  {journal} {Phys. Rev. Lett.}\ }\textbf {\bibinfo {volume} {103}},\ \bibinfo
  {pages} {057002} (\bibinfo {year} {2009})}\BibitemShut {NoStop}%
\bibitem [{\citenamefont {Bao}\ \emph {et~al.}(2009)\citenamefont {Bao} \emph
  {et~al.}}]{Bao09-Fe(TeSe)}%
  \BibitemOpen
  \bibfield  {author} {\bibinfo {author} {\bibfnamefont {W.}~\bibnamefont
  {Bao}} \emph {et~al.},\ }\href@noop {} {\bibfield  {journal} {\bibinfo
  {journal} {Phys. Rev. Lett.}\ }\textbf {\bibinfo {volume} {102}},\ \bibinfo
  {pages} {247001} (\bibinfo {year} {2009})}\BibitemShut {NoStop}%
\bibitem [{\citenamefont {Li}\ \emph {et~al.}(2009{\natexlab{a}})\citenamefont
  {Li}, \citenamefont {de~la Cruz}, \citenamefont {Huang}, \citenamefont
  {Chen}, \citenamefont {Lynn}, \citenamefont {Hu}, \citenamefont {Huang},
  \citenamefont {Hsu}, \citenamefont {Yeh}, \citenamefont {Wu},\ and\
  \citenamefont {Dai}}]{Li09-FeTe-1stOrderTransition}%
  \BibitemOpen
  \bibfield  {author} {\bibinfo {author} {\bibfnamefont {S.}~\bibnamefont
  {Li}}, \bibinfo {author} {\bibfnamefont {C.}~\bibnamefont {de~la Cruz}},
  \bibinfo {author} {\bibfnamefont {Q.}~\bibnamefont {Huang}}, \bibinfo
  {author} {\bibfnamefont {Y.}~\bibnamefont {Chen}}, \bibinfo {author}
  {\bibfnamefont {J.~W.}\ \bibnamefont {Lynn}}, \bibinfo {author}
  {\bibfnamefont {J.}~\bibnamefont {Hu}}, \bibinfo {author} {\bibfnamefont
  {Y.-L.}\ \bibnamefont {Huang}}, \bibinfo {author} {\bibfnamefont {F.-C.}\
  \bibnamefont {Hsu}}, \bibinfo {author} {\bibfnamefont {K.-W.}\ \bibnamefont
  {Yeh}}, \bibinfo {author} {\bibfnamefont {M.-K.}\ \bibnamefont {Wu}}, \ and\
  \bibinfo {author} {\bibfnamefont {P.}~\bibnamefont {Dai}},\ }\href@noop {}
  {\bibfield  {journal} {\bibinfo  {journal} {Phys. Rev. B}\ }\textbf {\bibinfo
  {volume} {79}},\ \bibinfo {pages} {054503} (\bibinfo {year}
  {2009}{\natexlab{a}})}\BibitemShut {NoStop}%
\bibitem [{\citenamefont {Yi}\ \emph {et~al.}(2013)\citenamefont {Yi},
  \citenamefont {Lu}, \citenamefont {Yu}, \citenamefont {Riggs}, \citenamefont
  {Chu}, \citenamefont {Lv}, \citenamefont {Liu}, \citenamefont {Lu},
  \citenamefont {Cui}, \citenamefont {Hashimoto}, \citenamefont {Mo},
  \citenamefont {Hussain}, \citenamefont {Chu}, \citenamefont {Fisher},
  \citenamefont {Si},\ and\ \citenamefont {Shen}}]{Yi13-OSMP-AxFe2-ySe2}%
  \BibitemOpen
  \bibfield  {author} {\bibinfo {author} {\bibfnamefont {M.}~\bibnamefont
  {Yi}}, \bibinfo {author} {\bibfnamefont {D.~H.}\ \bibnamefont {Lu}}, \bibinfo
  {author} {\bibfnamefont {R.}~\bibnamefont {Yu}}, \bibinfo {author}
  {\bibfnamefont {S.~C.}\ \bibnamefont {Riggs}}, \bibinfo {author}
  {\bibfnamefont {J.-H.}\ \bibnamefont {Chu}}, \bibinfo {author} {\bibfnamefont
  {B.}~\bibnamefont {Lv}}, \bibinfo {author} {\bibfnamefont {Z.~K.}\
  \bibnamefont {Liu}}, \bibinfo {author} {\bibfnamefont {M.}~\bibnamefont
  {Lu}}, \bibinfo {author} {\bibfnamefont {Y.-T.}\ \bibnamefont {Cui}},
  \bibinfo {author} {\bibfnamefont {M.}~\bibnamefont {Hashimoto}}, \bibinfo
  {author} {\bibfnamefont {S.-K.}\ \bibnamefont {Mo}}, \bibinfo {author}
  {\bibfnamefont {Z.}~\bibnamefont {Hussain}}, \bibinfo {author} {\bibfnamefont
  {C.~W.}\ \bibnamefont {Chu}}, \bibinfo {author} {\bibfnamefont {I.~R.}\
  \bibnamefont {Fisher}}, \bibinfo {author} {\bibfnamefont {Q.}~\bibnamefont
  {Si}}, \ and\ \bibinfo {author} {\bibfnamefont {Z.-X.}\ \bibnamefont
  {Shen}},\ }\href@noop {} {\bibfield  {journal} {\bibinfo  {journal} {Phys.
  Rev. Lett.}\ }\textbf {\bibinfo {volume} {110}},\ \bibinfo {pages} {067003}
  (\bibinfo {year} {2013})}\BibitemShut {NoStop}%
\bibitem [{\citenamefont {Gao}\ \emph {et~al.}(2014)\citenamefont {Gao},
  \citenamefont {Yu}, \citenamefont {Sun}, \citenamefont {Wang}, \citenamefont
  {Wang}, \citenamefont {Wu}, \citenamefont {Fang}, \citenamefont {Chen},
  \citenamefont {Guo}, \citenamefont {Zhang}, \citenamefont {Gu}, \citenamefont
  {Tian}, \citenamefont {Li}, \citenamefont {Liu}, \citenamefont {Li},
  \citenamefont {Li}, \citenamefont {Jiang}, \citenamefont {Yang},
  \citenamefont {Li}, \citenamefont {Si},\ and\ \citenamefont
  {Zhao}}]{Gao14-OSMP-245Phase-HighPressure}%
  \BibitemOpen
  \bibfield  {author} {\bibinfo {author} {\bibfnamefont {P.}~\bibnamefont
  {Gao}}, \bibinfo {author} {\bibfnamefont {R.}~\bibnamefont {Yu}}, \bibinfo
  {author} {\bibfnamefont {L.}~\bibnamefont {Sun}}, \bibinfo {author}
  {\bibfnamefont {H.}~\bibnamefont {Wang}}, \bibinfo {author} {\bibfnamefont
  {Z.}~\bibnamefont {Wang}}, \bibinfo {author} {\bibfnamefont {Q.}~\bibnamefont
  {Wu}}, \bibinfo {author} {\bibfnamefont {M.}~\bibnamefont {Fang}}, \bibinfo
  {author} {\bibfnamefont {G.}~\bibnamefont {Chen}}, \bibinfo {author}
  {\bibfnamefont {J.}~\bibnamefont {Guo}}, \bibinfo {author} {\bibfnamefont
  {C.}~\bibnamefont {Zhang}}, \bibinfo {author} {\bibfnamefont
  {D.}~\bibnamefont {Gu}}, \bibinfo {author} {\bibfnamefont {H.}~\bibnamefont
  {Tian}}, \bibinfo {author} {\bibfnamefont {J.}~\bibnamefont {Li}}, \bibinfo
  {author} {\bibfnamefont {J.}~\bibnamefont {Liu}}, \bibinfo {author}
  {\bibfnamefont {Y.}~\bibnamefont {Li}}, \bibinfo {author} {\bibfnamefont
  {X.}~\bibnamefont {Li}}, \bibinfo {author} {\bibfnamefont {S.}~\bibnamefont
  {Jiang}}, \bibinfo {author} {\bibfnamefont {K.}~\bibnamefont {Yang}},
  \bibinfo {author} {\bibfnamefont {A.}~\bibnamefont {Li}}, \bibinfo {author}
  {\bibfnamefont {Q.}~\bibnamefont {Si}}, \ and\ \bibinfo {author}
  {\bibfnamefont {Z.}~\bibnamefont {Zhao}},\ }\href@noop {} {\bibfield
  {journal} {\bibinfo  {journal} {Phys. Rev. B}\ }\textbf {\bibinfo {volume}
  {89}},\ \bibinfo {pages} {094514} (\bibinfo {year} {2014})}\BibitemShut
  {NoStop}%
\bibitem [{\citenamefont {Altshuler}\ and\ \citenamefont
  {Aronov}(1983)}]{Altshuler83-Localization-e-e}%
  \BibitemOpen
  \bibfield  {author} {\bibinfo {author} {\bibfnamefont {B.~L.}\ \bibnamefont
  {Altshuler}}\ and\ \bibinfo {author} {\bibfnamefont {A.~G.}\ \bibnamefont
  {Aronov}},\ }\href@noop {} {\bibfield  {journal} {\bibinfo  {journal} {Solid
  State Commun}\ }\textbf {\bibinfo {volume} {46}},\ \bibinfo {pages} {429}
  (\bibinfo {year} {1983})}\BibitemShut {NoStop}%
\bibitem [{\citenamefont {Yu}\ and\ \citenamefont
  {Si}(2013)}]{Yu13-OSMP-KxFe2-yK2}%
  \BibitemOpen
  \bibfield  {author} {\bibinfo {author} {\bibfnamefont {R.}~\bibnamefont
  {Yu}}\ and\ \bibinfo {author} {\bibfnamefont {Q.}~\bibnamefont {Si}},\
  }\href@noop {} {\bibfield  {journal} {\bibinfo  {journal} {Phys. Rev. Lett.}\
  }\textbf {\bibinfo {volume} {110}},\ \bibinfo {pages} {146402} (\bibinfo
  {year} {2013})}\BibitemShut {NoStop}%
\bibitem [{\citenamefont {Craco}\ \emph {et~al.}(2011)\citenamefont {Craco},
  \citenamefont {Laad},\ and\ \citenamefont {Leoni}}]{Craco11-Mott-KxFe2-ySe2}%
  \BibitemOpen
  \bibfield  {author} {\bibinfo {author} {\bibfnamefont {L.}~\bibnamefont
  {Craco}}, \bibinfo {author} {\bibfnamefont {M.~S.}\ \bibnamefont {Laad}}, \
  and\ \bibinfo {author} {\bibfnamefont {S.}~\bibnamefont {Leoni}},\
  }\href@noop {} {\bibfield  {journal} {\bibinfo  {journal} {Phys. Rev. B}\
  }\textbf {\bibinfo {volume} {84}},\ \bibinfo {pages} {224520} (\bibinfo
  {year} {2011})}\BibitemShut {NoStop}%
\bibitem [{\citenamefont {Yin}\ \emph {et~al.}(2012)\citenamefont {Yin},
  \citenamefont {Haule},\ and\ \citenamefont
  {Kotliar}}]{Yin12-powerLaw-Fe-Chalcogenides}%
  \BibitemOpen
  \bibfield  {author} {\bibinfo {author} {\bibfnamefont {Z.~P.}\ \bibnamefont
  {Yin}}, \bibinfo {author} {\bibfnamefont {K.}~\bibnamefont {Haule}}, \ and\
  \bibinfo {author} {\bibfnamefont {G.}~\bibnamefont {Kotliar}},\ }\href@noop
  {} {\bibfield  {journal} {\bibinfo  {journal} {Phys. Rev. B}\ }\textbf
  {\bibinfo {volume} {86}},\ \bibinfo {pages} {195141} (\bibinfo {year}
  {2012})}\BibitemShut {NoStop}%
\bibitem [{\citenamefont {Bascones}\ \emph {et~al.}(2012)\citenamefont
  {Bascones}, \citenamefont {Valenzuela},\ and\ \citenamefont
  {Calder\'on}}]{Bascones12-OSMP-Magnetic-Fe-SUCs}%
  \BibitemOpen
  \bibfield  {author} {\bibinfo {author} {\bibfnamefont {E.}~\bibnamefont
  {Bascones}}, \bibinfo {author} {\bibfnamefont {B.}~\bibnamefont
  {Valenzuela}}, \ and\ \bibinfo {author} {\bibfnamefont {M.~J.}\ \bibnamefont
  {Calder\'on}},\ }\href@noop {} {\bibfield  {journal} {\bibinfo  {journal}
  {Phys. Rev. B}\ }\textbf {\bibinfo {volume} {86}},\ \bibinfo {pages} {174508}
  (\bibinfo {year} {2012})}\BibitemShut {NoStop}%
\bibitem [{\citenamefont {Mott}(1987)}]{Mott87-MobilityEdge}%
  \BibitemOpen
  \bibfield  {author} {\bibinfo {author} {\bibfnamefont {N.~F.}\ \bibnamefont
  {Mott}},\ }\href@noop {} {\bibfield  {journal} {\bibinfo  {journal} {J
  Physics C: Solid State Physics}\ }\textbf {\bibinfo {volume} {20}},\ \bibinfo
  {pages} {3075} (\bibinfo {year} {1987})}\BibitemShut {NoStop}%
\bibitem [{\citenamefont {Basko}\ \emph {et~al.}(2006)\citenamefont {Basko},
  \citenamefont {Aleiner},\ and\ \citenamefont {Altshuler}}]{Basko06-MIT}%
  \BibitemOpen
  \bibfield  {author} {\bibinfo {author} {\bibfnamefont {D.}~\bibnamefont
  {Basko}}, \bibinfo {author} {\bibfnamefont {I.}~\bibnamefont {Aleiner}}, \
  and\ \bibinfo {author} {\bibfnamefont {B.}~\bibnamefont {Altshuler}},\
  }\href@noop {} {\bibfield  {journal} {\bibinfo  {journal} {Annals of
  Physics}\ }\textbf {\bibinfo {volume} {321}},\ \bibinfo {pages} {1126}
  (\bibinfo {year} {2006})}\BibitemShut {NoStop}%
\bibitem [{\citenamefont {Liu}\ \emph {et~al.}(2009)\citenamefont {Liu} \emph
  {et~al.}}]{Liu09-FeExcess-FeTeSe}%
  \BibitemOpen
  \bibfield  {author} {\bibinfo {author} {\bibfnamefont {T.~J.}\ \bibnamefont
  {Liu}} \emph {et~al.},\ }\href@noop {} {\bibfield  {journal} {\bibinfo
  {journal} {Phys. Rev. B}\ }\textbf {\bibinfo {volume} {80}},\ \bibinfo
  {pages} {174509} (\bibinfo {year} {2009})}\BibitemShut {NoStop}%
\bibitem [{\citenamefont {Chang}\ \emph {et~al.}(2012)\citenamefont {Chang},
  \citenamefont {Luo}, \citenamefont {Wu}, \citenamefont {Hsu}, \citenamefont
  {Huang}, \citenamefont {Wu}, \citenamefont {Wu},\ and\ \citenamefont
  {Wang}}]{Chang12-FeTe1-xSex-WeakLocalization}%
  \BibitemOpen
  \bibfield  {author} {\bibinfo {author} {\bibfnamefont {H.~H.}\ \bibnamefont
  {Chang}}, \bibinfo {author} {\bibfnamefont {J.~Y.}\ \bibnamefont {Luo}},
  \bibinfo {author} {\bibfnamefont {C.~T.}\ \bibnamefont {Wu}}, \bibinfo
  {author} {\bibfnamefont {F.~C.}\ \bibnamefont {Hsu}}, \bibinfo {author}
  {\bibfnamefont {T.~W.}\ \bibnamefont {Huang}}, \bibinfo {author}
  {\bibfnamefont {P.~M.}\ \bibnamefont {Wu}}, \bibinfo {author} {\bibfnamefont
  {M.~K.}\ \bibnamefont {Wu}}, \ and\ \bibinfo {author} {\bibfnamefont {M.~J.}\
  \bibnamefont {Wang}},\ }\href@noop {} {\bibfield  {journal} {\bibinfo
  {journal} {Supercond. Sci. Technol.}\ }\textbf {\bibinfo {volume} {25}},\
  \bibinfo {pages} {035004} (\bibinfo {year} {2012})}\BibitemShut {NoStop}%
\bibitem [{\citenamefont {Ando}\ \emph {et~al.}(2004)\citenamefont {Ando},
  \citenamefont {Komiya}, \citenamefont {Segawa}, \citenamefont {Ono},\ and\
  \citenamefont {Kurita}}]{Ando04-PhaseDiagram-HTC-Resistivity}%
  \BibitemOpen
  \bibfield  {author} {\bibinfo {author} {\bibfnamefont {Y.}~\bibnamefont
  {Ando}}, \bibinfo {author} {\bibfnamefont {S.}~\bibnamefont {Komiya}},
  \bibinfo {author} {\bibfnamefont {K.}~\bibnamefont {Segawa}}, \bibinfo
  {author} {\bibfnamefont {S.}~\bibnamefont {Ono}}, \ and\ \bibinfo {author}
  {\bibfnamefont {Y.}~\bibnamefont {Kurita}},\ }\href@noop {} {\bibfield
  {journal} {\bibinfo  {journal} {Phys. Rev. Lett.}\ }\textbf {\bibinfo
  {volume} {93}},\ \bibinfo {pages} {267001} (\bibinfo {year}
  {2004})}\BibitemShut {NoStop}%
\bibitem [{\citenamefont {Song}\ \emph {et~al.}(2011)\citenamefont {Song},
  \citenamefont {Hong}, \citenamefont {Min}, \citenamefont {Kwon},
  \citenamefont {Lee}, \citenamefont {Jung},\ and\ \citenamefont
  {Rhyee}}]{Song11-FeSe-SUC-Normal-State}%
  \BibitemOpen
  \bibfield  {author} {\bibinfo {author} {\bibfnamefont {Y.~J.}\ \bibnamefont
  {Song}}, \bibinfo {author} {\bibfnamefont {J.~B.}\ \bibnamefont {Hong}},
  \bibinfo {author} {\bibfnamefont {B.~H.}\ \bibnamefont {Min}}, \bibinfo
  {author} {\bibfnamefont {Y.~S.}\ \bibnamefont {Kwon}}, \bibinfo {author}
  {\bibfnamefont {K.~J.}\ \bibnamefont {Lee}}, \bibinfo {author} {\bibfnamefont
  {M.~H.}\ \bibnamefont {Jung}}, \ and\ \bibinfo {author} {\bibfnamefont
  {J.-S.}\ \bibnamefont {Rhyee}},\ }\href@noop {} {\bibfield  {journal}
  {\bibinfo  {journal} {J. Korean Phys.Soc.}\ }\textbf {\bibinfo {volume}
  {59}},\ \bibinfo {pages} {312} (\bibinfo {year} {2011})}\BibitemShut
  {NoStop}%
\bibitem [{\citenamefont {Liu}\ \emph {et~al.}(2011)\citenamefont {Liu},
  \citenamefont {Kremer},\ and\ \citenamefont {Lin}}]{Liu11-Fe1+dTe1-xSex}%
  \BibitemOpen
  \bibfield  {author} {\bibinfo {author} {\bibfnamefont {Y.}~\bibnamefont
  {Liu}}, \bibinfo {author} {\bibfnamefont {R.~K.}\ \bibnamefont {Kremer}}, \
  and\ \bibinfo {author} {\bibfnamefont {C.~T.}\ \bibnamefont {Lin}},\
  }\href@noop {} {\bibfield  {journal} {\bibinfo  {journal} {Supercond. Sci.
  Technol.}\ }\textbf {\bibinfo {volume} {24}},\ \bibinfo {pages} {035012}
  (\bibinfo {year} {2011})}\BibitemShut {NoStop}%
\bibitem [{\citenamefont {Yadav}\ and\ \citenamefont
  {Paulose}(2010)}]{Yadav10-Feexcess-Fe(TeSe)}%
  \BibitemOpen
  \bibfield  {author} {\bibinfo {author} {\bibfnamefont {C.~S.}\ \bibnamefont
  {Yadav}}\ and\ \bibinfo {author} {\bibfnamefont {P.~L.}\ \bibnamefont
  {Paulose}},\ }\href@noop {} {\bibfield  {journal} {\bibinfo  {journal} {J.
  Appl. Phys.}\ }\textbf {\bibinfo {volume} {107}},\ \bibinfo {pages} {083908}
  (\bibinfo {year} {2010})}\BibitemShut {NoStop}%
\bibitem [{\citenamefont {Deguchi}\ \emph {et~al.}(2014)\citenamefont
  {Deguchi}, \citenamefont {Demura}, \citenamefont {Yamaki}, \citenamefont
  {Hara}, \citenamefont {Denholme}, \citenamefont {Fujioka}, \citenamefont
  {Okazaki}, \citenamefont {Takeya},\ and\ \citenamefont
  {Takano}}]{Deguchi13-FeTeSe-beverage}%
  \BibitemOpen
  \bibfield  {author} {\bibinfo {author} {\bibfnamefont {K.}~\bibnamefont
  {Deguchi}}, \bibinfo {author} {\bibfnamefont {S.}~\bibnamefont {Demura}},
  \bibinfo {author} {\bibfnamefont {T.}~\bibnamefont {Yamaki}}, \bibinfo
  {author} {\bibfnamefont {H.}~\bibnamefont {Hara}}, \bibinfo {author}
  {\bibfnamefont {S.}~\bibnamefont {Denholme}}, \bibinfo {author}
  {\bibfnamefont {M.}~\bibnamefont {Fujioka}}, \bibinfo {author} {\bibfnamefont
  {H.}~\bibnamefont {Okazaki}}, \bibinfo {author} {\bibfnamefont
  {H.}~\bibnamefont {Takeya}}, \ and\ \bibinfo {author} {\bibfnamefont
  {T.~Y.~Y.}\ \bibnamefont {Takano}},\ }\href@noop {} {\bibfield  {journal}
  {\bibinfo  {journal} {J Supercond Nov Magn}\ }\textbf {\bibinfo {volume}
  {27}},\ \bibinfo {pages} {305} (\bibinfo {year} {2014})}\BibitemShut
  {NoStop}%
\bibitem [{\citenamefont {Sun}\ \emph {et~al.}(2014)\citenamefont {Sun},
  \citenamefont {Tsuchiya}, \citenamefont {Taen}, \citenamefont {Yamada},
  \citenamefont {Pyon}, \citenamefont {Shi},\ and\ \citenamefont
  {Tamegai}}]{Sun2014-O-FeTeSe}%
  \BibitemOpen
  \bibfield  {author} {\bibinfo {author} {\bibfnamefont {Y.}~\bibnamefont
  {Sun}}, \bibinfo {author} {\bibfnamefont {Y.}~\bibnamefont {Tsuchiya}},
  \bibinfo {author} {\bibfnamefont {T.}~\bibnamefont {Taen}}, \bibinfo {author}
  {\bibfnamefont {T.}~\bibnamefont {Yamada}}, \bibinfo {author} {\bibfnamefont
  {S.}~\bibnamefont {Pyon}}, \bibinfo {author} {\bibfnamefont {A.~S. T. E.~Z.}\
  \bibnamefont {Shi}}, \ and\ \bibinfo {author} {\bibfnamefont
  {T.}~\bibnamefont {Tamegai}},\ }\href@noop {} {\bibfield  {journal} {\bibinfo
   {journal} {Scientific Reports}\ }\textbf {\bibinfo {volume} {4}},\ \bibinfo
  {pages} {4585} (\bibinfo {year} {2014})}\BibitemShut {NoStop}%
\bibitem [{\citenamefont {Pallecchi}\ \emph {et~al.}(2011)\citenamefont
  {Pallecchi}, \citenamefont {Tropeano}, \citenamefont {Lamura}, \citenamefont
  {Martinelli}, \citenamefont {Palenzona},\ and\ \citenamefont
  {Putti}}]{Pallecchi11-NormalState-Fe11-Fe111}%
  \BibitemOpen
  \bibfield  {author} {\bibinfo {author} {\bibfnamefont {I.}~\bibnamefont
  {Pallecchi}}, \bibinfo {author} {\bibfnamefont {M.}~\bibnamefont {Tropeano}},
  \bibinfo {author} {\bibfnamefont {C.~F.~G.}\ \bibnamefont {Lamura}}, \bibinfo
  {author} {\bibfnamefont {A.}~\bibnamefont {Martinelli}}, \bibinfo {author}
  {\bibfnamefont {A.}~\bibnamefont {Palenzona}}, \ and\ \bibinfo {author}
  {\bibfnamefont {M.}~\bibnamefont {Putti}},\ }\href@noop {} {\bibfield
  {journal} {\bibinfo  {journal} {J Supercond Nov Mag}\ }\textbf {\bibinfo
  {volume} {24}},\ \bibinfo {pages} {1751} (\bibinfo {year}
  {2011})}\BibitemShut {NoStop}%
\bibitem [{\citenamefont {Martinelli}\ \emph {et~al.}(2010)\citenamefont
  {Martinelli}, \citenamefont {Palenzona}, \citenamefont {Tropeano},
  \citenamefont {Ferdeghini}, \citenamefont {Putti}, \citenamefont {Cimberle},
  \citenamefont {Nguyen}, \citenamefont {Affronte},\ and\ \citenamefont
  {Ritter}}]{Martinelli10-Fe(TeSe)-PhaseDiagram}%
  \BibitemOpen
  \bibfield  {author} {\bibinfo {author} {\bibfnamefont {A.}~\bibnamefont
  {Martinelli}}, \bibinfo {author} {\bibfnamefont {A.}~\bibnamefont
  {Palenzona}}, \bibinfo {author} {\bibfnamefont {M.}~\bibnamefont {Tropeano}},
  \bibinfo {author} {\bibfnamefont {C.}~\bibnamefont {Ferdeghini}}, \bibinfo
  {author} {\bibfnamefont {M.}~\bibnamefont {Putti}}, \bibinfo {author}
  {\bibfnamefont {M.~R.}\ \bibnamefont {Cimberle}}, \bibinfo {author}
  {\bibfnamefont {T.~D.}\ \bibnamefont {Nguyen}}, \bibinfo {author}
  {\bibfnamefont {M.}~\bibnamefont {Affronte}}, \ and\ \bibinfo {author}
  {\bibfnamefont {C.}~\bibnamefont {Ritter}},\ }\href@noop {} {\bibfield
  {journal} {\bibinfo  {journal} {Phys. Rev. B}\ }\textbf {\bibinfo {volume}
  {81}},\ \bibinfo {pages} {094115} (\bibinfo {year} {2010})}\BibitemShut
  {NoStop}%
\bibitem [{\citenamefont {Okada}\ \emph {et~al.}(2009)\citenamefont {Okada},
  \citenamefont {Takahashi}, \citenamefont {Mizuguchi}, \citenamefont
  {Takano},\ and\ \citenamefont {Takahashi}}]{Okada09-PhaseTrans-FeTe}%
  \BibitemOpen
  \bibfield  {author} {\bibinfo {author} {\bibfnamefont {H.}~\bibnamefont
  {Okada}}, \bibinfo {author} {\bibfnamefont {H.}~\bibnamefont {Takahashi}},
  \bibinfo {author} {\bibfnamefont {Y.}~\bibnamefont {Mizuguchi}}, \bibinfo
  {author} {\bibfnamefont {Y.}~\bibnamefont {Takano}}, \ and\ \bibinfo {author}
  {\bibfnamefont {H.}~\bibnamefont {Takahashi}},\ }\href@noop {} {\bibfield
  {journal} {\bibinfo  {journal} {J. Phys. Soc. Jpn.}\ }\textbf {\bibinfo
  {volume} {78}},\ \bibinfo {pages} {083709} (\bibinfo {year}
  {2009})}\BibitemShut {NoStop}%
\bibitem [{\citenamefont {Mizuguchi}\ \emph {et~al.}(2008)\citenamefont
  {Mizuguchi}, \citenamefont {Tomioka}, \citenamefont {Tsuda}, \citenamefont
  {Yamaguchi},\ and\ \citenamefont {Takano}}]{Mizuguchi08-FeSe-27K-pressure}%
  \BibitemOpen
  \bibfield  {author} {\bibinfo {author} {\bibfnamefont {Y.}~\bibnamefont
  {Mizuguchi}}, \bibinfo {author} {\bibfnamefont {F.}~\bibnamefont {Tomioka}},
  \bibinfo {author} {\bibfnamefont {S.}~\bibnamefont {Tsuda}}, \bibinfo
  {author} {\bibfnamefont {T.}~\bibnamefont {Yamaguchi}}, \ and\ \bibinfo
  {author} {\bibfnamefont {Y.}~\bibnamefont {Takano}},\ }\href@noop {}
  {\bibfield  {journal} {\bibinfo  {journal} {Appl. Phys. Lett.}\ }\textbf
  {\bibinfo {volume} {93}},\ \bibinfo {pages} {152505} (\bibinfo {year}
  {2008})}\BibitemShut {NoStop}%
\bibitem [{\citenamefont {Margadonna}\ \emph {et~al.}(2009)\citenamefont
  {Margadonna}, \citenamefont {Takabayashi}, \citenamefont {Ohishi},
  \citenamefont {Mizuguchi}, \citenamefont {Takano}, \citenamefont {Kagayama},
  \citenamefont {Nakagawa}, \citenamefont {Takata},\ and\ \citenamefont
  {Prassides}}]{Margadonna09-FeSe-Pressure}%
  \BibitemOpen
  \bibfield  {author} {\bibinfo {author} {\bibfnamefont {S.}~\bibnamefont
  {Margadonna}}, \bibinfo {author} {\bibfnamefont {Y.}~\bibnamefont
  {Takabayashi}}, \bibinfo {author} {\bibfnamefont {Y.}~\bibnamefont {Ohishi}},
  \bibinfo {author} {\bibfnamefont {Y.}~\bibnamefont {Mizuguchi}}, \bibinfo
  {author} {\bibfnamefont {Y.}~\bibnamefont {Takano}}, \bibinfo {author}
  {\bibfnamefont {T.}~\bibnamefont {Kagayama}}, \bibinfo {author}
  {\bibfnamefont {T.}~\bibnamefont {Nakagawa}}, \bibinfo {author}
  {\bibfnamefont {M.}~\bibnamefont {Takata}}, \ and\ \bibinfo {author}
  {\bibfnamefont {K.}~\bibnamefont {Prassides}},\ }\href@noop {} {\bibfield
  {journal} {\bibinfo  {journal} {Phys. Rev. B}\ }\textbf {\bibinfo {volume}
  {80}},\ \bibinfo {pages} {064506} (\bibinfo {year} {2009})}\BibitemShut
  {NoStop}%
\bibitem [{\citenamefont {Imai}\ \emph {et~al.}(2009)\citenamefont {Imai},
  \citenamefont {Ahilan}, \citenamefont {Ning}, \citenamefont {McQueen},\ and\
  \citenamefont {Cava}}]{Imai09-FeSe-HighPressure}%
  \BibitemOpen
  \bibfield  {author} {\bibinfo {author} {\bibfnamefont {T.}~\bibnamefont
  {Imai}}, \bibinfo {author} {\bibfnamefont {K.}~\bibnamefont {Ahilan}},
  \bibinfo {author} {\bibfnamefont {F.~L.}\ \bibnamefont {Ning}}, \bibinfo
  {author} {\bibfnamefont {T.~M.}\ \bibnamefont {McQueen}}, \ and\ \bibinfo
  {author} {\bibfnamefont {R.~J.}\ \bibnamefont {Cava}},\ }\href@noop {}
  {\bibfield  {journal} {\bibinfo  {journal} {Phys. Rev. Lett.}\ }\textbf
  {\bibinfo {volume} {102}},\ \bibinfo {pages} {177005} (\bibinfo {year}
  {2009})}\BibitemShut {NoStop}%
\bibitem [{\citenamefont {Medvedev}\ \emph {et~al.}(2009)\citenamefont
  {Medvedev}, \citenamefont {McQueen}, \citenamefont {Troyan}, \citenamefont
  {Palasyuk}, \citenamefont {Eremets}, \citenamefont {Cava}, \citenamefont
  {Naghavi}, \citenamefont {Casper}, \citenamefont {Ksenofontov}, \citenamefont
  {Wortmann},\ and\ \citenamefont {Felser}}]{Medvedev09-FeSe-Pressure}%
  \BibitemOpen
  \bibfield  {author} {\bibinfo {author} {\bibfnamefont {S.}~\bibnamefont
  {Medvedev}}, \bibinfo {author} {\bibfnamefont {T.~M.}\ \bibnamefont
  {McQueen}}, \bibinfo {author} {\bibfnamefont {I.~A.}\ \bibnamefont {Troyan}},
  \bibinfo {author} {\bibfnamefont {T.}~\bibnamefont {Palasyuk}}, \bibinfo
  {author} {\bibfnamefont {M.~I.}\ \bibnamefont {Eremets}}, \bibinfo {author}
  {\bibfnamefont {R.~J.}\ \bibnamefont {Cava}}, \bibinfo {author}
  {\bibfnamefont {S.}~\bibnamefont {Naghavi}}, \bibinfo {author} {\bibfnamefont
  {F.}~\bibnamefont {Casper}}, \bibinfo {author} {\bibfnamefont
  {V.}~\bibnamefont {Ksenofontov}}, \bibinfo {author} {\bibfnamefont
  {G.}~\bibnamefont {Wortmann}}, \ and\ \bibinfo {author} {\bibfnamefont
  {C.}~\bibnamefont {Felser}},\ }\href@noop {} {\bibfield  {journal} {\bibinfo
  {journal} {Nat. Mater.}\ }\textbf {\bibinfo {volume} {8}},\ \bibinfo {pages}
  {630} (\bibinfo {year} {2009})}\BibitemShut {NoStop}%
\bibitem [{\citenamefont {Li}\ \emph {et~al.}(2009{\natexlab{b}})\citenamefont
  {Li}, \citenamefont {Yang}, \citenamefont {Ge}, \citenamefont {Pi},
  \citenamefont {Xu}, \citenamefont {Wang}, \citenamefont {Sun},\ and\
  \citenamefont {Zhang}}]{Li09-FeSex-Pressure}%
  \BibitemOpen
  \bibfield  {author} {\bibinfo {author} {\bibfnamefont {L.}~\bibnamefont
  {Li}}, \bibinfo {author} {\bibfnamefont {Z.}~\bibnamefont {Yang}}, \bibinfo
  {author} {\bibfnamefont {M.}~\bibnamefont {Ge}}, \bibinfo {author}
  {\bibfnamefont {L.}~\bibnamefont {Pi}}, \bibinfo {author} {\bibfnamefont
  {J.}~\bibnamefont {Xu}}, \bibinfo {author} {\bibfnamefont {B.}~\bibnamefont
  {Wang}}, \bibinfo {author} {\bibfnamefont {Y.}~\bibnamefont {Sun}}, \ and\
  \bibinfo {author} {\bibfnamefont {Y.}~\bibnamefont {Zhang}},\ }\href@noop {}
  {\bibfield  {journal} {\bibinfo  {journal} {J Supercond Novel Mag}\ }\textbf
  {\bibinfo {volume} {22}},\ \bibinfo {pages} {667} (\bibinfo {year}
  {2009}{\natexlab{b}})}\BibitemShut {NoStop}%
\bibitem [{\citenamefont {Braithwaite}\ \emph {et~al.}(2009)\citenamefont
  {Braithwaite}, \citenamefont {Salce}, \citenamefont {Lapertot}, \citenamefont
  {Bourdarot}, \citenamefont {Martin}, \citenamefont {Aoki}, ,\ and\
  \citenamefont {Hanflan}}]{Braithwaite09-FeSe-HighPressure}%
  \BibitemOpen
  \bibfield  {author} {\bibinfo {author} {\bibfnamefont {D.}~\bibnamefont
  {Braithwaite}}, \bibinfo {author} {\bibfnamefont {B.}~\bibnamefont {Salce}},
  \bibinfo {author} {\bibfnamefont {G.}~\bibnamefont {Lapertot}}, \bibinfo
  {author} {\bibfnamefont {F.}~\bibnamefont {Bourdarot}}, \bibinfo {author}
  {\bibfnamefont {C.}~\bibnamefont {Martin}}, \bibinfo {author} {\bibfnamefont
  {D.}~\bibnamefont {Aoki}}, , \ and\ \bibinfo {author} {\bibfnamefont
  {M.}~\bibnamefont {Hanflan}},\ }\href@noop {} {\bibfield  {journal} {\bibinfo
   {journal} {J. Phys: Condens. Matter}\ }\textbf {\bibinfo {volume} {21}},\
  \bibinfo {pages} {232202} (\bibinfo {year} {2009})}\BibitemShut {NoStop}%
\bibitem [{\citenamefont {Garbarino}\ \emph {et~al.}(2009)\citenamefont
  {Garbarino}, \citenamefont {Sow}, \citenamefont {Lejay}, \citenamefont
  {Sulpice}, \citenamefont {Toulemonde}, \citenamefont {Mezouar},\ and\
  \citenamefont {Nunez-Regueiro}}]{Garbarino09-FeSe-HighPressure}%
  \BibitemOpen
  \bibfield  {author} {\bibinfo {author} {\bibfnamefont {G.}~\bibnamefont
  {Garbarino}}, \bibinfo {author} {\bibfnamefont {A.}~\bibnamefont {Sow}},
  \bibinfo {author} {\bibfnamefont {P.}~\bibnamefont {Lejay}}, \bibinfo
  {author} {\bibfnamefont {A.}~\bibnamefont {Sulpice}}, \bibinfo {author}
  {\bibfnamefont {P.}~\bibnamefont {Toulemonde}}, \bibinfo {author}
  {\bibfnamefont {M.}~\bibnamefont {Mezouar}}, \ and\ \bibinfo {author}
  {\bibfnamefont {M.}~\bibnamefont {Nunez-Regueiro}},\ }\href@noop {}
  {\bibfield  {journal} {\bibinfo  {journal} {Europhys. Lett.}\ }\textbf
  {\bibinfo {volume} {86}},\ \bibinfo {pages} {27001} (\bibinfo {year}
  {2009})}\BibitemShut {NoStop}%
\bibitem [{\citenamefont {Horigane}\ \emph {et~al.}(2009)\citenamefont
  {Horigane}, \citenamefont {Takeshita}, \citenamefont {Lee}, \citenamefont
  {Hiraka},\ and\ \citenamefont {Yamada}}]{Horigane09-FeTe05Se05}%
  \BibitemOpen
  \bibfield  {author} {\bibinfo {author} {\bibfnamefont {K.}~\bibnamefont
  {Horigane}}, \bibinfo {author} {\bibfnamefont {N.}~\bibnamefont {Takeshita}},
  \bibinfo {author} {\bibfnamefont {C.-H.}\ \bibnamefont {Lee}}, \bibinfo
  {author} {\bibfnamefont {H.}~\bibnamefont {Hiraka}}, \ and\ \bibinfo {author}
  {\bibfnamefont {K.}~\bibnamefont {Yamada}},\ }\href@noop {} {\bibfield
  {journal} {\bibinfo  {journal} {J. Phys. Soc. Jpn.}\ }\textbf {\bibinfo
  {volume} {78}},\ \bibinfo {pages} {063705} (\bibinfo {year}
  {2009})}\BibitemShut {NoStop}%
\bibitem [{\citenamefont {Tsoi}\ \emph {et~al.}(2009)\citenamefont {Tsoi},
  \citenamefont {Stemshorn}, \citenamefont {Vohra}, \citenamefont {Wu},
  \citenamefont {Hsu}, \citenamefont {Huang}, \citenamefont {Wu}, \citenamefont
  {Yeh},\ and\ \citenamefont {Weir}}]{Tsoi09-FeSe05Te05-HighPressure}%
  \BibitemOpen
  \bibfield  {author} {\bibinfo {author} {\bibfnamefont {G.}~\bibnamefont
  {Tsoi}}, \bibinfo {author} {\bibfnamefont {A.}~\bibnamefont {Stemshorn}},
  \bibinfo {author} {\bibfnamefont {K.}~\bibnamefont {Vohra}}, \bibinfo
  {author} {\bibfnamefont {P.}~\bibnamefont {Wu}}, \bibinfo {author}
  {\bibfnamefont {F.}~\bibnamefont {Hsu}}, \bibinfo {author} {\bibfnamefont
  {Y.}~\bibnamefont {Huang}}, \bibinfo {author} {\bibfnamefont
  {M.}~\bibnamefont {Wu}}, \bibinfo {author} {\bibfnamefont {K.}~\bibnamefont
  {Yeh}}, \ and\ \bibinfo {author} {\bibfnamefont {S.}~\bibnamefont {Weir}},\
  }\href@noop {} {\bibfield  {journal} {\bibinfo  {journal} {J. Phys: Condens.
  Matter}\ }\textbf {\bibinfo {volume} {21}},\ \bibinfo {pages} {232201}
  (\bibinfo {year} {2009})}\BibitemShut {NoStop}%
\bibitem [{\citenamefont {Huang}\ \emph {et~al.}(2009)\citenamefont {Huang},
  \citenamefont {Chou}, \citenamefont {Tseng}, \citenamefont {Huang},
  \citenamefont {Hsu}, \citenamefont {Yeh}, \citenamefont {Wu},\ and\
  \citenamefont {Yang}}]{Huang09-FeSe088-FeSe05Te05-Pressure}%
  \BibitemOpen
  \bibfield  {author} {\bibinfo {author} {\bibfnamefont {C.-L.}\ \bibnamefont
  {Huang}}, \bibinfo {author} {\bibfnamefont {C.-C.}\ \bibnamefont {Chou}},
  \bibinfo {author} {\bibfnamefont {K.-F.}\ \bibnamefont {Tseng}}, \bibinfo
  {author} {\bibfnamefont {Y.-L.}\ \bibnamefont {Huang}}, \bibinfo {author}
  {\bibfnamefont {F.-C.}\ \bibnamefont {Hsu}}, \bibinfo {author} {\bibfnamefont
  {K.-W.}\ \bibnamefont {Yeh}}, \bibinfo {author} {\bibfnamefont {M.-K.}\
  \bibnamefont {Wu}}, \ and\ \bibinfo {author} {\bibfnamefont {H.-D.}\
  \bibnamefont {Yang}},\ }\href@noop {} {\bibfield  {journal} {\bibinfo
  {journal} {J. Phys. Soc. Jpn.}\ }\textbf {\bibinfo {volume} {78}},\ \bibinfo
  {pages} {084710} (\bibinfo {year} {2009})}\BibitemShut {NoStop}%
\bibitem [{\citenamefont {Stemshorn}\ \emph {et~al.}(2010)\citenamefont
  {Stemshorn} \emph {et~al.}}]{Stemshorn10-FeSe-HighPressure}%
  \BibitemOpen
  \bibfield  {author} {\bibinfo {author} {\bibfnamefont {A.~K.}\ \bibnamefont
  {Stemshorn}} \emph {et~al.},\ }\href@noop {} {\bibfield  {journal} {\bibinfo
  {journal} {J. Mater. Res.}\ }\textbf {\bibinfo {volume} {25}},\ \bibinfo
  {pages} {396} (\bibinfo {year} {2010})}\BibitemShut {NoStop}%
\bibitem [{\citenamefont {Fang}\ \emph {et~al.}(2008)\citenamefont {Fang},
  \citenamefont {Pham}, \citenamefont {Qian}, \citenamefont {Liu},
  \citenamefont {Vehstedt}, \citenamefont {Liu}, \citenamefont {Spinu},\ and\
  \citenamefont {Mao}}]{Fang08-Fe(SeTe)}%
  \BibitemOpen
  \bibfield  {author} {\bibinfo {author} {\bibfnamefont {M.~H.}\ \bibnamefont
  {Fang}}, \bibinfo {author} {\bibfnamefont {H.~M.}\ \bibnamefont {Pham}},
  \bibinfo {author} {\bibfnamefont {B.}~\bibnamefont {Qian}}, \bibinfo {author}
  {\bibfnamefont {T.~J.}\ \bibnamefont {Liu}}, \bibinfo {author} {\bibfnamefont
  {E.~K.}\ \bibnamefont {Vehstedt}}, \bibinfo {author} {\bibfnamefont
  {Y.}~\bibnamefont {Liu}}, \bibinfo {author} {\bibfnamefont {L.}~\bibnamefont
  {Spinu}}, \ and\ \bibinfo {author} {\bibfnamefont {Z.~Q.}\ \bibnamefont
  {Mao}},\ }\href@noop {} {\bibfield  {journal} {\bibinfo  {journal} {Phys.
  Rev. B}\ }\textbf {\bibinfo {volume} {78}},\ \bibinfo {pages} {224503}
  (\bibinfo {year} {2008})}\BibitemShut {NoStop}%
\bibitem [{\citenamefont {Dong}\ \emph {et~al.}(2011)\citenamefont {Dong},
  \citenamefont {Wang}, \citenamefont {Li}, \citenamefont {Chen}, \citenamefont
  {Yuan},\ and\ \citenamefont {Fang}}]{Dong11-PhaseDiagram-FeTeSe}%
  \BibitemOpen
  \bibfield  {author} {\bibinfo {author} {\bibfnamefont {C.}~\bibnamefont
  {Dong}}, \bibinfo {author} {\bibfnamefont {H.}~\bibnamefont {Wang}}, \bibinfo
  {author} {\bibfnamefont {Z.}~\bibnamefont {Li}}, \bibinfo {author}
  {\bibfnamefont {J.}~\bibnamefont {Chen}}, \bibinfo {author} {\bibfnamefont
  {H.~Q.}\ \bibnamefont {Yuan}}, \ and\ \bibinfo {author} {\bibfnamefont
  {M.}~\bibnamefont {Fang}},\ }\href@noop {} {\bibfield  {journal} {\bibinfo
  {journal} {Phys. Rev. B}\ }\textbf {\bibinfo {volume} {84}},\ \bibinfo
  {pages} {224506} (\bibinfo {year} {2011})}\BibitemShut {NoStop}%
\bibitem [{\citenamefont {Kawasaki}\ \emph {et~al.}(2012)\citenamefont
  {Kawasaki}, \citenamefont {Deguchi}, \citenamefont {Demura}, \citenamefont
  {Watanabe}, \citenamefont {Okazaki}, \citenamefont {Ozaki}, \citenamefont
  {Yamaguchi}, \citenamefont {Takeya},\ and\ \citenamefont
  {Takano}}]{Kawasaki12-Fe(TeSe)-O-anneal-PhaseDiagram}%
  \BibitemOpen
  \bibfield  {author} {\bibinfo {author} {\bibfnamefont {Y.}~\bibnamefont
  {Kawasaki}}, \bibinfo {author} {\bibfnamefont {K.}~\bibnamefont {Deguchi}},
  \bibinfo {author} {\bibfnamefont {S.}~\bibnamefont {Demura}}, \bibinfo
  {author} {\bibfnamefont {T.}~\bibnamefont {Watanabe}}, \bibinfo {author}
  {\bibfnamefont {H.}~\bibnamefont {Okazaki}}, \bibinfo {author} {\bibfnamefont
  {T.}~\bibnamefont {Ozaki}}, \bibinfo {author} {\bibfnamefont
  {T.}~\bibnamefont {Yamaguchi}}, \bibinfo {author} {\bibfnamefont
  {H.}~\bibnamefont {Takeya}}, \ and\ \bibinfo {author} {\bibfnamefont
  {Y.}~\bibnamefont {Takano}},\ }\href@noop {} {\bibfield  {journal} {\bibinfo
  {journal} {Solid State Commun.}\ }\textbf {\bibinfo {volume} {152}},\
  \bibinfo {pages} {1135} (\bibinfo {year} {2012})}\BibitemShut {NoStop}%
\bibitem [{\citenamefont {Machida}\ \emph
  {et~al.}(2013{\natexlab{a}})\citenamefont {Machida}, \citenamefont {Kogure},
  \citenamefont {Kato}, \citenamefont {Nakamura}, \citenamefont {Takeya},
  \citenamefont {Mochiku}, \citenamefont {Ooi}, \citenamefont {Mizuguchi},
  \citenamefont {Takano}, \citenamefont {Hirata},\ and\ \citenamefont
  {Sakata}}]{Machida13-FeTe-STM}%
  \BibitemOpen
  \bibfield  {author} {\bibinfo {author} {\bibfnamefont {T.}~\bibnamefont
  {Machida}}, \bibinfo {author} {\bibfnamefont {K.}~\bibnamefont {Kogure}},
  \bibinfo {author} {\bibfnamefont {T.}~\bibnamefont {Kato}}, \bibinfo {author}
  {\bibfnamefont {H.}~\bibnamefont {Nakamura}}, \bibinfo {author}
  {\bibfnamefont {H.}~\bibnamefont {Takeya}}, \bibinfo {author} {\bibfnamefont
  {T.}~\bibnamefont {Mochiku}}, \bibinfo {author} {\bibfnamefont
  {S.}~\bibnamefont {Ooi}}, \bibinfo {author} {\bibfnamefont {Y.}~\bibnamefont
  {Mizuguchi}}, \bibinfo {author} {\bibfnamefont {Y.}~\bibnamefont {Takano}},
  \bibinfo {author} {\bibfnamefont {K.}~\bibnamefont {Hirata}}, \ and\ \bibinfo
  {author} {\bibfnamefont {H.}~\bibnamefont {Sakata}},\ }\href@noop {}
  {\bibfield  {journal} {\bibinfo  {journal} {Phys. Rev. B}\ }\textbf {\bibinfo
  {volume} {87}},\ \bibinfo {pages} {214508} (\bibinfo {year}
  {2013}{\natexlab{a}})}\BibitemShut {NoStop}%
\bibitem [{\citenamefont {Machida}\ \emph
  {et~al.}(2013{\natexlab{b}})\citenamefont {Machida}, \citenamefont
  {Morohoshi}, \citenamefont {Takimoto}, \citenamefont {Nakamura},
  \citenamefont {Takeya}, \citenamefont {Mochiku}, \citenamefont {Ooi},
  \citenamefont {Mizuguchi}, \citenamefont {Takano}, \citenamefont {Hirata},\
  and\ \citenamefont {Sakata}}]{Machida13-FeTe-excess-Fe}%
  \BibitemOpen
  \bibfield  {author} {\bibinfo {author} {\bibfnamefont {T.}~\bibnamefont
  {Machida}}, \bibinfo {author} {\bibfnamefont {D.}~\bibnamefont {Morohoshi}},
  \bibinfo {author} {\bibfnamefont {K.}~\bibnamefont {Takimoto}}, \bibinfo
  {author} {\bibfnamefont {H.}~\bibnamefont {Nakamura}}, \bibinfo {author}
  {\bibfnamefont {H.}~\bibnamefont {Takeya}}, \bibinfo {author} {\bibfnamefont
  {T.}~\bibnamefont {Mochiku}}, \bibinfo {author} {\bibfnamefont
  {S.}~\bibnamefont {Ooi}}, \bibinfo {author} {\bibfnamefont {Y.}~\bibnamefont
  {Mizuguchi}}, \bibinfo {author} {\bibfnamefont {Y.}~\bibnamefont {Takano}},
  \bibinfo {author} {\bibfnamefont {K.}~\bibnamefont {Hirata}}, \ and\ \bibinfo
  {author} {\bibfnamefont {H.}~\bibnamefont {Sakata}},\ }\href@noop {}
  {\bibfield  {journal} {\bibinfo  {journal} {Physica C}\ }\textbf {\bibinfo
  {volume} {484}},\ \bibinfo {pages} {19} (\bibinfo {year}
  {2013}{\natexlab{b}})}\BibitemShut {NoStop}%
\bibitem [{\citenamefont {Hu}\ \emph {et~al.}(2013)\citenamefont {Hu},
  \citenamefont {Liu}, \citenamefont {Qian},\ and\ \citenamefont
  {Mao}}]{Hu13-Fe1+y(Te1-xSex)}%
  \BibitemOpen
  \bibfield  {author} {\bibinfo {author} {\bibfnamefont {J.}~\bibnamefont
  {Hu}}, \bibinfo {author} {\bibfnamefont {T.~J.}\ \bibnamefont {Liu}},
  \bibinfo {author} {\bibfnamefont {B.}~\bibnamefont {Qian}}, \ and\ \bibinfo
  {author} {\bibfnamefont {Z.~Q.}\ \bibnamefont {Mao}},\ }\href@noop {}
  {\bibfield  {journal} {\bibinfo  {journal} {Phys. Rev. B}\ }\textbf {\bibinfo
  {volume} {88}},\ \bibinfo {pages} {094505} (\bibinfo {year}
  {2013})}\BibitemShut {NoStop}%
\bibitem [{\citenamefont {Cao}\ \emph {et~al.}(2011)\citenamefont {Cao},
  \citenamefont {Shen}, \citenamefont {Chen}, \citenamefont {Yuan},
  \citenamefont {Kang},\ and\ \citenamefont {Zhang}}]{Cao11-FeExcess-FeTeSe}%
  \BibitemOpen
  \bibfield  {author} {\bibinfo {author} {\bibfnamefont {S.}~\bibnamefont
  {Cao}}, \bibinfo {author} {\bibfnamefont {S.}~\bibnamefont {Shen}}, \bibinfo
  {author} {\bibfnamefont {L.}~\bibnamefont {Chen}}, \bibinfo {author}
  {\bibfnamefont {S.}~\bibnamefont {Yuan}}, \bibinfo {author} {\bibfnamefont
  {B.}~\bibnamefont {Kang}}, \ and\ \bibinfo {author} {\bibfnamefont
  {J.}~\bibnamefont {Zhang}},\ }\href@noop {} {\bibfield  {journal} {\bibinfo
  {journal} {J. Appl. Phys.}\ }\textbf {\bibinfo {volume} {110}},\ \bibinfo
  {pages} {033914} (\bibinfo {year} {2011})}\BibitemShut {NoStop}%
\bibitem [{\citenamefont {Taraphder}\ and\ \citenamefont
  {Coleman}(1991)}]{Taraphder91-HF-negative-U-Anderson-Model}%
  \BibitemOpen
  \bibfield  {author} {\bibinfo {author} {\bibfnamefont {A.}~\bibnamefont
  {Taraphder}}\ and\ \bibinfo {author} {\bibfnamefont {P.}~\bibnamefont
  {Coleman}},\ }\href@noop {} {\bibfield  {journal} {\bibinfo  {journal} {Phys.
  Rev. Lett.}\ }\textbf {\bibinfo {volume} {66}},\ \bibinfo {pages} {2814}
  (\bibinfo {year} {1991})}\BibitemShut {NoStop}%
\bibitem [{\citenamefont {Ralph}\ and\ \citenamefont
  {Buhrman}(1992)}]{Ralph92-Kondo-No-MagImpurity}%
  \BibitemOpen
  \bibfield  {author} {\bibinfo {author} {\bibfnamefont {D.~C.}\ \bibnamefont
  {Ralph}}\ and\ \bibinfo {author} {\bibfnamefont {R.~A.}\ \bibnamefont
  {Buhrman}},\ }\href@noop {} {\bibfield  {journal} {\bibinfo  {journal} {Phys.
  Rev. Lett.}\ }\textbf {\bibinfo {volume} {69}},\ \bibinfo {pages} {2118}
  (\bibinfo {year} {1992})}\BibitemShut {NoStop}%
\bibitem [{\citenamefont {Anderson}\ \emph {et~al.}(1979)\citenamefont
  {Anderson}, \citenamefont {Abrahams},\ and\ \citenamefont
  {Ramakrishnan}}]{Anderson79-NonlinearConductance}%
  \BibitemOpen
  \bibfield  {author} {\bibinfo {author} {\bibfnamefont {P.~W.}\ \bibnamefont
  {Anderson}}, \bibinfo {author} {\bibfnamefont {E.}~\bibnamefont {Abrahams}},
  \ and\ \bibinfo {author} {\bibfnamefont {T.~V.}\ \bibnamefont
  {Ramakrishnan}},\ }\href@noop {} {\bibfield  {journal} {\bibinfo  {journal}
  {Phys. Rev. Lett.}\ }\textbf {\bibinfo {volume} {43}},\ \bibinfo {pages}
  {718} (\bibinfo {year} {1979})}\BibitemShut {NoStop}%
\bibitem [{\citenamefont {Dolan}\ and\ \citenamefont
  {Osheroff}(1979)}]{Dolan79-nonlinearConductance}%
  \BibitemOpen
  \bibfield  {author} {\bibinfo {author} {\bibfnamefont {G.~J.}\ \bibnamefont
  {Dolan}}\ and\ \bibinfo {author} {\bibfnamefont {D.~D.}\ \bibnamefont
  {Osheroff}},\ }\href@noop {} {\bibfield  {journal} {\bibinfo  {journal}
  {Phys. Rev. Lett.}\ }\textbf {\bibinfo {volume} {43}},\ \bibinfo {pages}
  {721} (\bibinfo {year} {1979})}\BibitemShut {NoStop}%
\bibitem [{\citenamefont {Sacepe}\ \emph {et~al.}(2010)\citenamefont {Sacepe},
  \citenamefont {Chapelier}, \citenamefont {Baturina}, \citenamefont
  {Vinokur},\ and\ \citenamefont {Sanquer}}]{Sacepe10-PseudoGap-TiN}%
  \BibitemOpen
  \bibfield  {author} {\bibinfo {author} {\bibfnamefont {B.}~\bibnamefont
  {Sacepe}}, \bibinfo {author} {\bibfnamefont {C.}~\bibnamefont {Chapelier}},
  \bibinfo {author} {\bibfnamefont {T.~I.}\ \bibnamefont {Baturina}}, \bibinfo
  {author} {\bibfnamefont {V.~M.}\ \bibnamefont {Vinokur}}, \ and\ \bibinfo
  {author} {\bibfnamefont {M.~R. B.~M.}\ \bibnamefont {Sanquer}},\ }\href@noop
  {} {\bibfield  {journal} {\bibinfo  {journal} {Nat Commun}\ }\textbf
  {\bibinfo {volume} {1}},\ \bibinfo {pages} {140} (\bibinfo {year}
  {2010})}\BibitemShut {NoStop}%
\bibitem [{\citenamefont {Lee}\ \emph {et~al.}(2006)\citenamefont {Lee},
  \citenamefont {Nagaosa},\ and\ \citenamefont
  {Wen}}]{Lee06-Doping-Mott-Insulator-Review}%
  \BibitemOpen
  \bibfield  {author} {\bibinfo {author} {\bibfnamefont {P.~A.}\ \bibnamefont
  {Lee}}, \bibinfo {author} {\bibfnamefont {N.}~\bibnamefont {Nagaosa}}, \ and\
  \bibinfo {author} {\bibfnamefont {X.-G.}\ \bibnamefont {Wen}},\ }\href@noop
  {} {\bibfield  {journal} {\bibinfo  {journal} {Rev. Mod. Phys.}\ }\textbf
  {\bibinfo {volume} {78}},\ \bibinfo {pages} {17} (\bibinfo {year}
  {2006})}\BibitemShut {NoStop}%
\bibitem [{\citenamefont {Ketterson}\ and\ \citenamefont
  {Song}(1999)}]{Ketterson-Song-SUC-textbook}%
  \BibitemOpen
  \bibfield  {author} {\bibinfo {author} {\bibfnamefont {J.~B.}\ \bibnamefont
  {Ketterson}}\ and\ \bibinfo {author} {\bibfnamefont {S.~N.}\ \bibnamefont
  {Song}},\ }\href@noop {} {\emph {\bibinfo {title} {Superconductivity}}}\
  (\bibinfo  {publisher} {Cambridge University Press},\ \bibinfo {year}
  {1999})\BibitemShut {NoStop}%
\bibitem [{\citenamefont {Williams}\ \emph {et~al.}(2009)\citenamefont
  {Williams}, \citenamefont {McQueen}, \citenamefont {Ksenofontov},
  \citenamefont {Felser},\ and\ \citenamefont
  {Cava}}]{Williams09-(FeCu)Se-MIT}%
  \BibitemOpen
  \bibfield  {author} {\bibinfo {author} {\bibfnamefont {A.~J.}\ \bibnamefont
  {Williams}}, \bibinfo {author} {\bibfnamefont {T.~M.}\ \bibnamefont
  {McQueen}}, \bibinfo {author} {\bibfnamefont {V.}~\bibnamefont
  {Ksenofontov}}, \bibinfo {author} {\bibfnamefont {C.}~\bibnamefont {Felser}},
  \ and\ \bibinfo {author} {\bibfnamefont {R.~J.}\ \bibnamefont {Cava}},\
  }\href@noop {} {\bibfield  {journal} {\bibinfo  {journal} {J. Phys.: Condens.
  Matter}\ }\textbf {\bibinfo {volume} {21}},\ \bibinfo {pages} {305701}
  (\bibinfo {year} {2009})}\BibitemShut {NoStop}%
\end{thebibliography}%

\end{document}